\documentclass[]{llncs}

\pdfoutput=1

\usepackage[margin=2.5cm]{geometry}

\usepackage{epsfig}
\usepackage{graphics}
\usepackage{graphicx}
\usepackage{amsmath}
\usepackage{amsfonts}
\usepackage{amssymb}
\usepackage{amsfonts}
\usepackage{url}
\usepackage{subfigure}
\usepackage{multicol}
\usepackage{captcont}
\usepackage{rotating}
\usepackage{xspace}
\usepackage{wasysym}

\newtheorem{finding}{Finding}

\newcommand{\h}{$\mathbf{H}$\xspace}
\newcommand{\p}{$\mathbf{P}$\xspace}

\begin{document}

\title{Are motorways rational from slime mould's point of view? }

\author{
Andrew~Adamatzky\inst{1}
\and
Selim~Akl\inst{2}
\and
Ramon~Alonso-Sanz\inst{3}
\and
Wesley~van~Dessel\inst{4}
\and
Zuwairie~Ibrahim\inst{5}
\and
Andrew~Ilachinski\inst{6}
\and
Jeff~Jones\inst{1}
\and
Anne~V.~D.~M.~Kayem\inst{7}
\and
Genaro~J.~Mart\'{i}nez\inst{8}
\and
Pedro~de~Oliveira\inst{9}
\and 
Mikhail~Prokopenko\inst{10}
\and
Theresa~Schubert\inst{11}
\and
Peter Sloot\inst{12}
\and
Emanuele~Strano\inst{13}
\and 
Xin-She~Yang\inst{14}
}

\institute{
University of the West of England, Bristol, UK
\and
Queen's University, Kingston, Ontario, Canada
\and
Universidad Politecnica de Madrid,  Madrid, Spain
\and
Scientific Institute of Public Health,  Brussels, Belgium
\and
Universiti Teknologi Malaysia, Johor Darul Takzim, Malaysia
\and
Center for Naval Analysis, Alexandria,  USA
\and
University of Cape Town, Cape Town, South Africa
\and
Universidad Nacional Aut\'onoma de M\'exico, M\'exico 
\and
Mackenzie Presbyterian University, S\~{a}o Paulo, Brazil
\and
CSIRO Information and Communication Technologies Centre, Sydney, Australia
\and
Bauhaus-Universit\"{a}t Weimar, Weimar, Germany
\and
University of Amsterdam, Amsterdam, The Netherlands
\and
EPFL, Lausanne, Switzerland 
\and
National Physical Laboratory, Teddington, United Kingdom
}

\maketitle

\begin{abstract}
We analyse the results of our experimental laboratory approximation of motorways networks  
with slime mould \emph{Physarum polycephalum}. Motorway networks of fourteen 
geographical areas are considered: Australia,  Africa, Belgium, Brazil, Canada, China, Germany, 
Iberia, Italy, Malaysia, Mexico, The Netherlands, UK, USA.  For each geographical entity we represented
major urban areas by oat flakes and inoculated the slime mould in a capital.  After
slime mould spanned all urban areas with a network of its protoplasmic tubes we 
extracted a generalised Physarum graph from the network and compared the graphs with 
an abstract motorway graph using most common measures. The measures employed are 
the number of independent cycles, cohesion, shortest paths lengths, diameter, 
the Harary index and the Randi\'{c} index. We obtained a series of intriguing results, and
found that the slime mould approximates best of all the motorway graphs of 
Belgium, Canada and China, and that for all entities studied the best 
match between Physarum and motorway graphs is detected by the Randi\'{c} index (molecular branching index).

\vspace{0.5cm}

\end{abstract}

\section{Introduction}

The increase of long-distance travel and subsequent reconfiguration of vehicular  and social networks~\cite{larsen}
requires novel and unconventional approaches towards analysis of dynamical processes in complex 
transport networks~\cite{barrat_2008},  routing and localisation of vehicular networks~\cite{Olariu},
optimisation of interactions between different parts of a transport network during scheduling of the road expansion and 
maintenance~\cite{taplin}, and shaping of transport network structure~\cite{beuthe}. ``The concept of a network is useful, but it may be misleading, because it gives the impression that it arises from a coherent planning process, whereas in practice almost all road networks have evolved gradually, being added to and altered over time to meet the ever-changing needs and demands of travellers. $\langle$\ldots$\rangle$ even our more recently built network of roads, the motorways designated for exclusive use of motor vehicles, has been result of nearly a century of thinking and rethinking about what should be built.''~\cite{davies}

Modern motorway networks are based on the millennium-long emergence of roads. First there were 
prehistoric trackways --- mesolithic footways established by early men 'seeking most direct and convenient alternatives by process
of trial and error' \cite{belloc_1924,davies}  or  based on the routes selected by the animals~\cite{taylor_1979}, timber trackways and 
droveways (used primarily by cattle). Further development of roads  was country-specific. Talking about England we can speculate that 
Romans were building roads along pre-existing trackways and possibly along Ridgeways. In the 1700s turnpikes were established. They were based  
on pre-existing roads with few local diversions~\cite{bogart_2007}. The turnpikes were then substituted by single carriageways, dual carriageways and, finally, motorways~\cite{davies}.  Thus backtracking the history of motorways we arrive to pathways developed by living creatures. 
How were the pathways developed? Were they efficient?  More or less accurate answers could be found by imitating the road network development with living substrates. 

While choosing a biological object to imitate the growth of road networks we want it to be experimental laboratory friendly, easy to cultivate and handle, 
and convenient to analyse. Ants would indeed be the first candidate. They do develop trails very similarly to pre-historic people. 
The great deal of impressive results has been published on ant-colony inspired computing~\cite{dorigo,solnon}. However ant colonies require 
substantial laboratory resources, experience and time in handling them. Actually very few, if any, papers were published on experimental laboratory implementation of ant-based optimisation (also there is the issue of what happens if ant collide on the paths, or fight),  the prevalent majority of publications being theoretical. There is however an object which is extremely easy to cultivate and handle, and which exhibits remarkably good foraging behaviour and development of intra-cellular transport networks. This is the plasmodium of \emph{Physarum polycephalum}. Plasmodium is a vegetative stage of acellular slime mould \emph{P. polycephalum}, 
 a single cell with many nuclei, which feeds on microscopic particles~\cite{stephenson_2000}. When foraging for its food the plasmodium propagates towards sources of food particles and microbes, surrounds them, secretes enzymes and digests the food.  Typically, the plasmodium forms a congregation of protoplasm covering the food source. When several sources of nutrients are scattered in the plasmodium's range, the plasmodium forms a network of protoplasmic tubes connecting the masses of protoplasm at the food sources. 
A structure of the protoplasmic networks is apparently optimal, in a sense that it covers all sources of nutrients and provides a robust and speedy transportation of nutrients and metabolites in the plasmodium's body~\cite{nakagaki_2001,nakagaki_2001a,nakagaki_iima_2007}.

Motorway networks are designed for efficient vehicular transportation of goods and passengers, protoplasmic networks are developed for efficient 
intra-cellular transportation of nutrients and metabolites. Is there a similarity between these two networks? 

To uncover analogies between biological and human-made transport networks and to project behavioural traits of biological networks onto development of vehicular transport networks we conducted a series of experimental laboratory studies on evaluation and approximation of motorway networks by \emph{P. polycephalum} in fourteen geographical regions:   Africa, Australia, Belgium, Brazil, Canada, China, Germany, Iberia, Italy, Malaysia, Mexico, The Netherlands, UK, and USA~\cite{PhysarumIberia}--\cite{PhysarumChina},\cite{PhysarumItaly}. We represented each region with an agar plate, imitated major urban areas with oat flakes, inoculated plasmodium of \emph{P. polycephalum} in a capital, and analysed structures of protoplasmic networks developed.
For all regions studied in laboratory experiments~\cite{PhysarumIberia}--\cite{PhysarumChina},\cite{PhysarumItaly}, we found that, 
the network of protoplasmic tubes grown by plasmodium matches, at least partly, the network of human-made transport arteries. The shape of a country and the exact spatial distribution of urban areas, represented by sources of nutrients, may play a key role in determining the exact structure of the plasmodium network. In the present paper we aim to answer two principal questions. What measures, apart from straightforward comparison of edges
between motorway and plasmodium networks, are reliable indicators of matching? Which regions have the most `Physarum friendly' motorway networks, i.e. show highest degree of matching between motorways and protoplasmic networks along several measures? In the course of investigation 
we got the answers to these questions and also obtained a few quite intriguing results on hierarchies of motorway and protoplasmic networks based on measures and topological indices.

\section{Experimental}

\begin{figure}[!tbp]
\centering
\subfigure[Africa]{\includegraphics[width=0.49\textwidth]{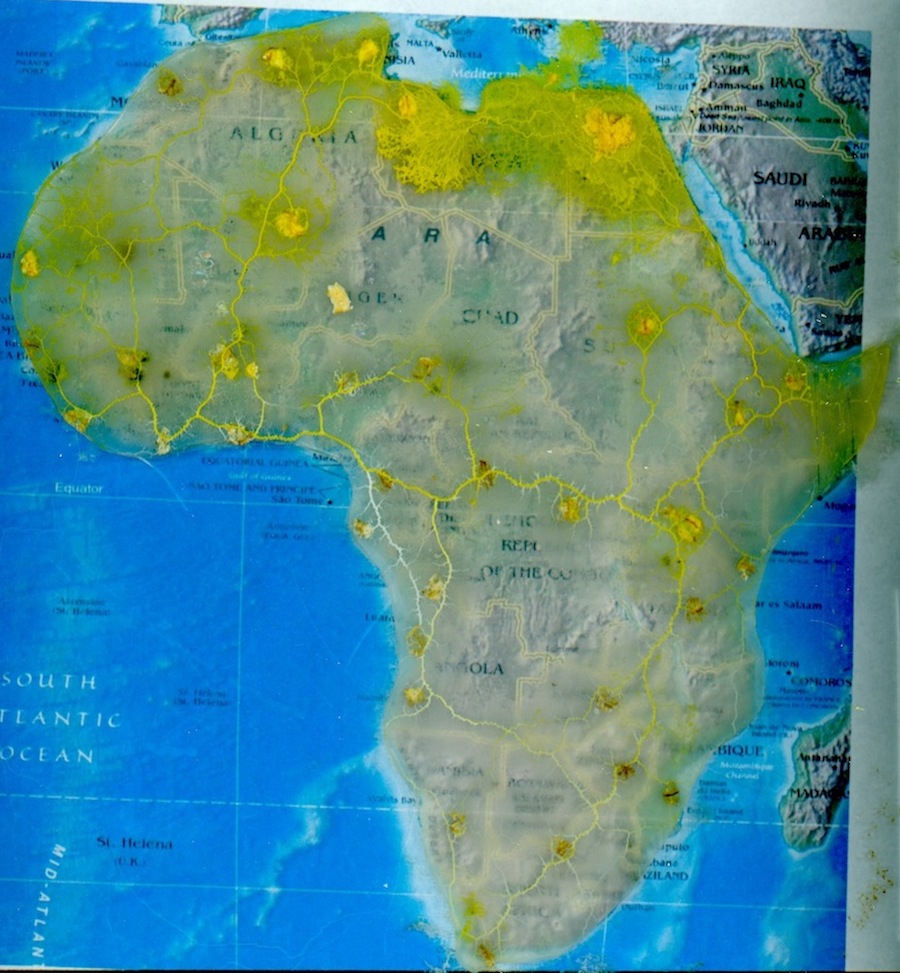}}
\subfigure[USA]{\includegraphics[width=0.49\textwidth]{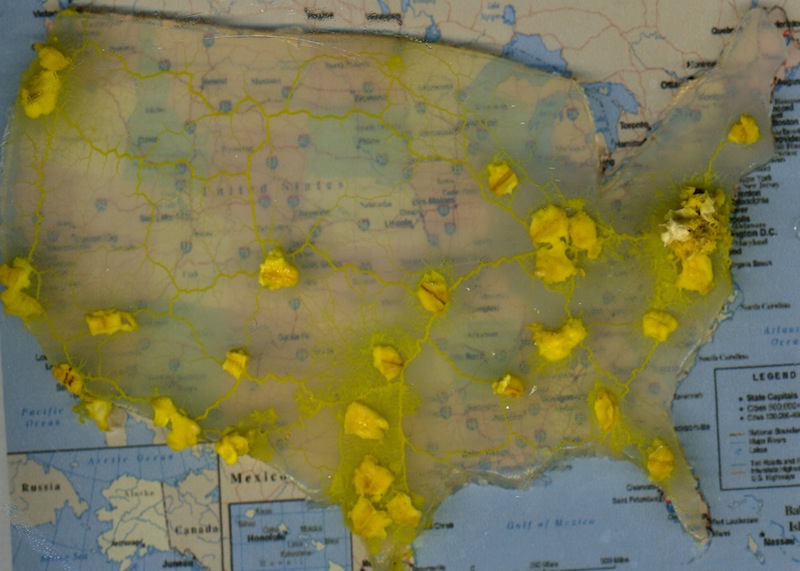}}
\caption{Experimental laboratory images of protoplasmic networks developed by slime mould \emph{P. polycephalum} on maps.}
\label{onmaps}
\end{figure}

\begin{figure}[!tbp]
\centering
\subfigure[Africa]{\includegraphics[width=0.25\textwidth]{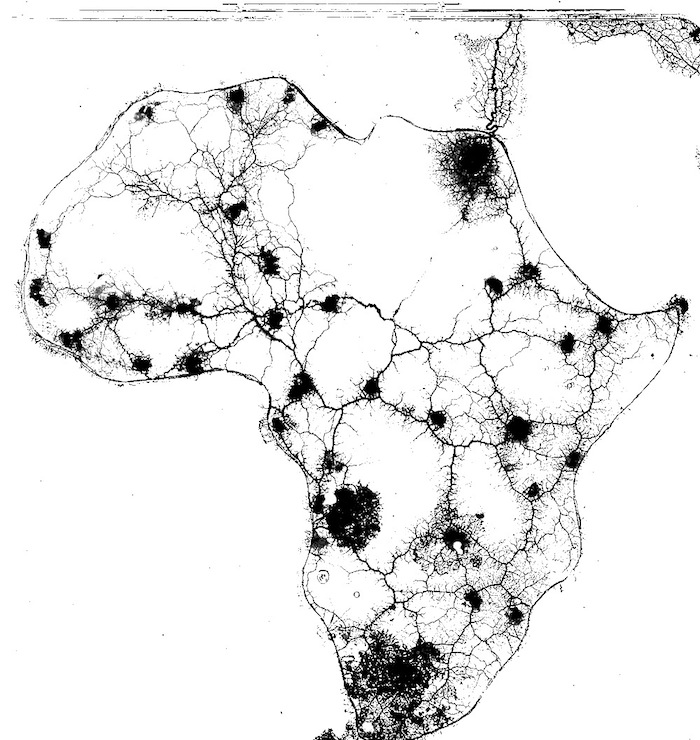}}
\subfigure[Australia]{\includegraphics[width=0.25\textwidth]{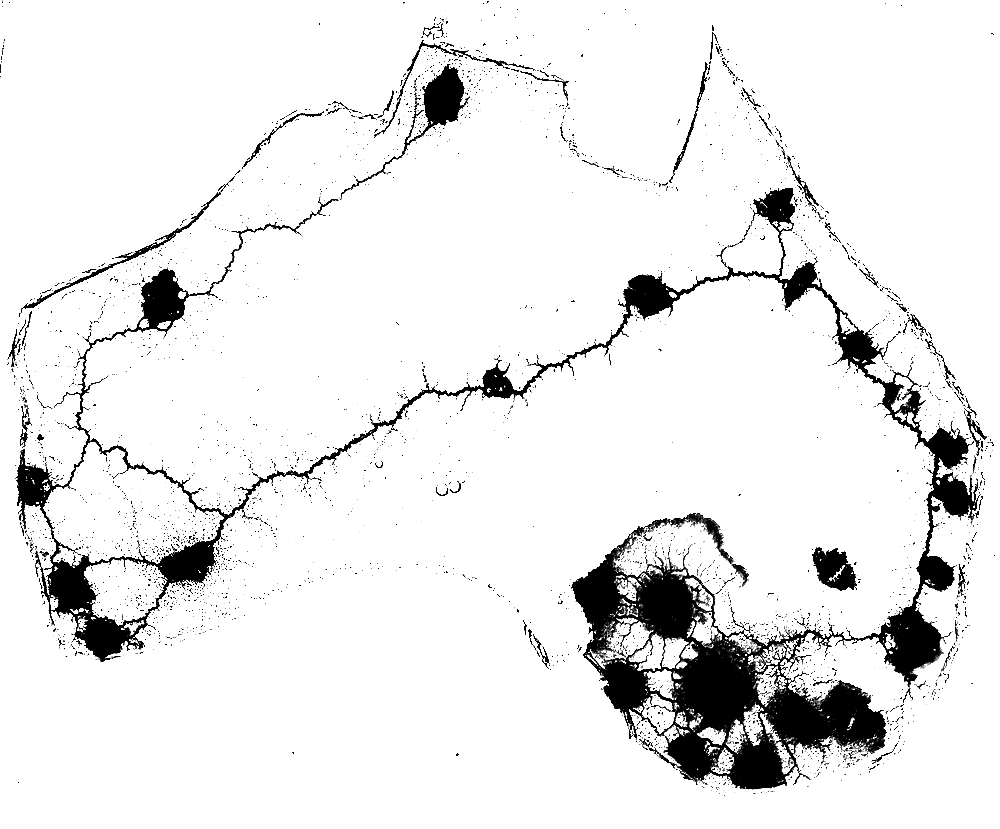}}
\subfigure[Belgium]{\includegraphics[width=0.25\textwidth]{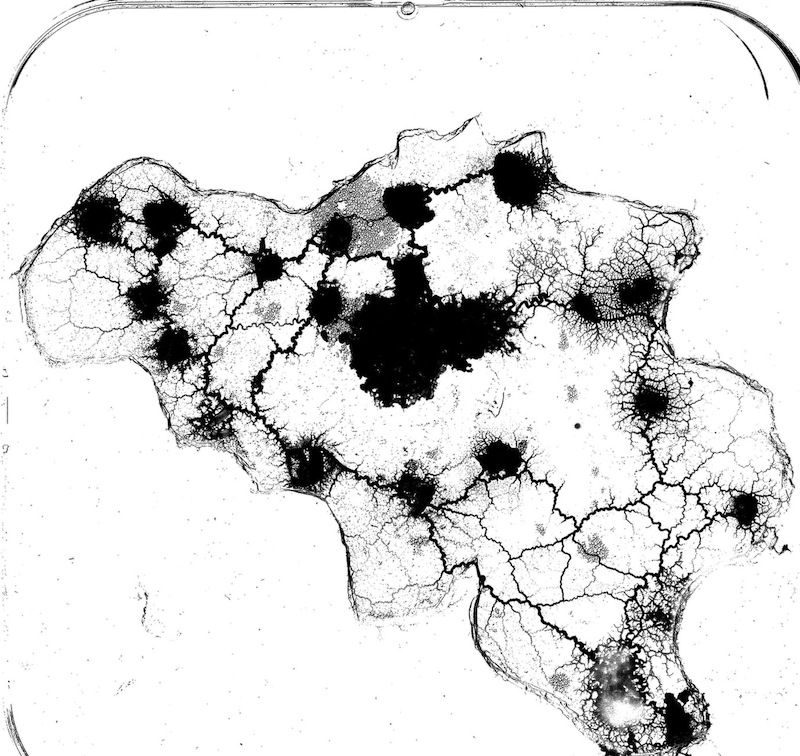}}
\subfigure[Brazil]{\includegraphics[width=0.25\textwidth]{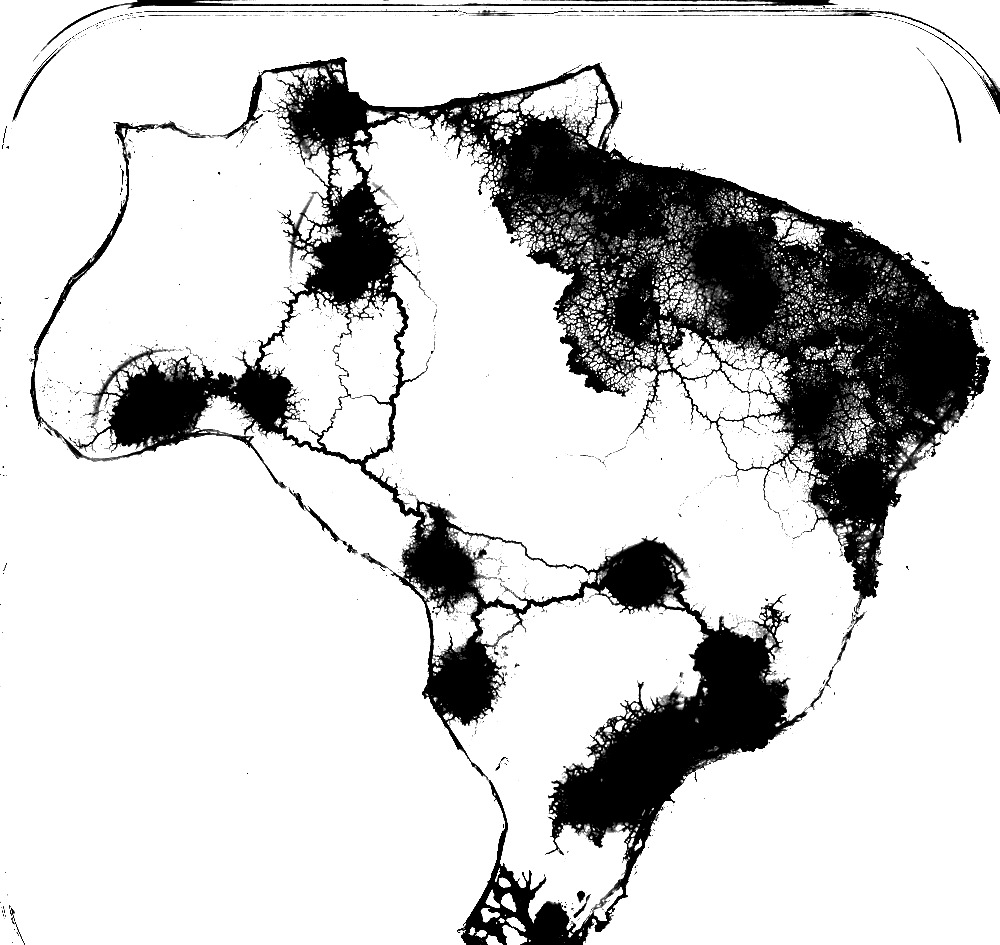}}
\subfigure[Canada]{\includegraphics[width=0.25\textwidth]{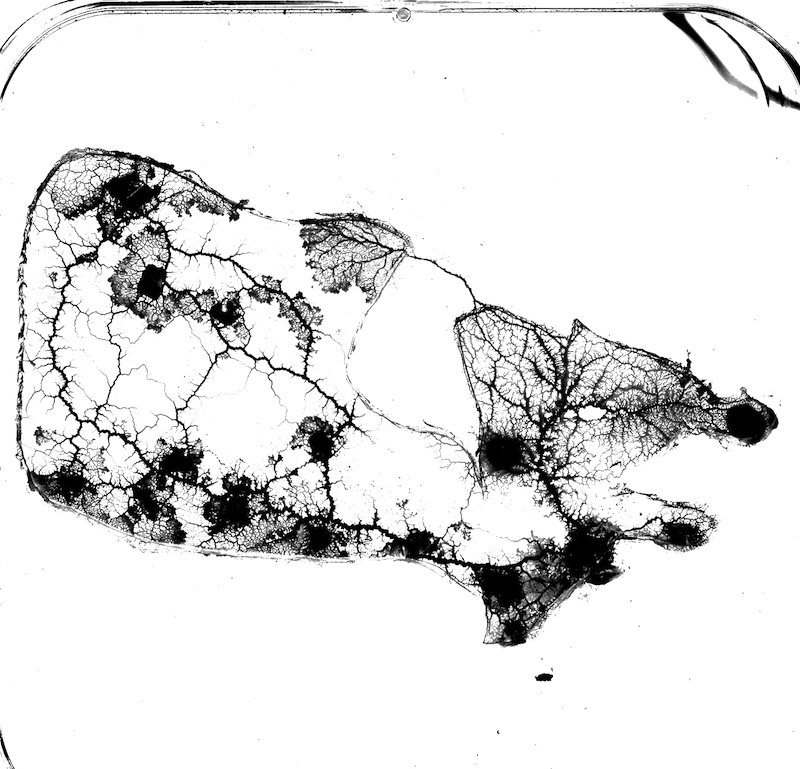}}
\subfigure[China]{\includegraphics[width=0.25\textwidth]{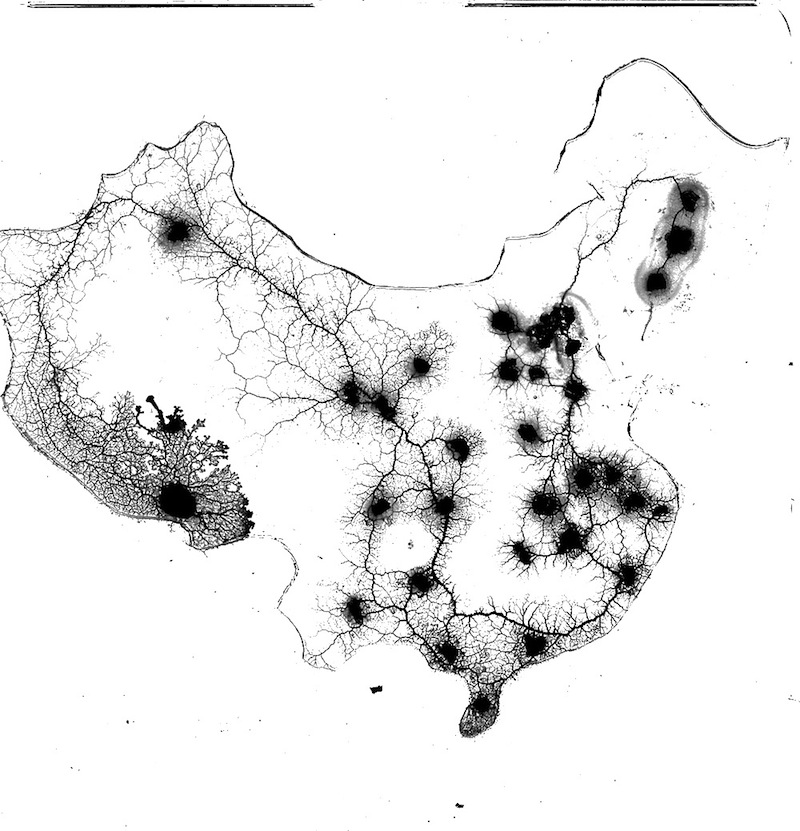}}
\subfigure[Germany]{\includegraphics[width=0.25\textwidth]{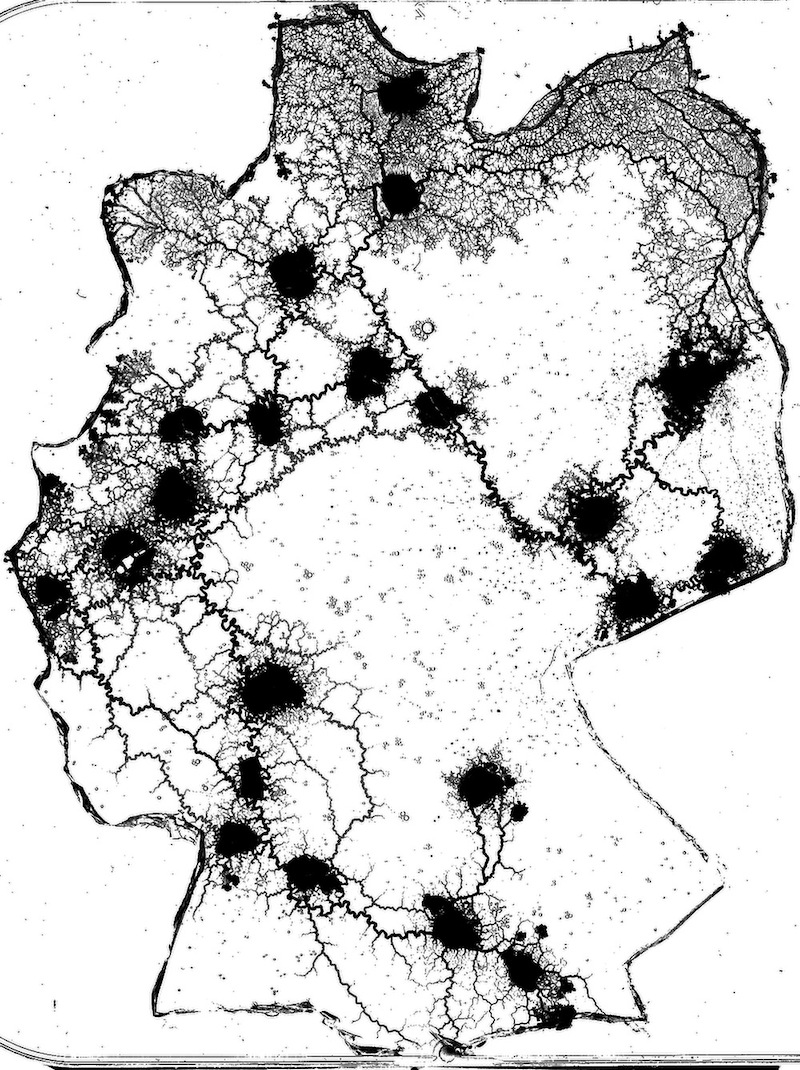}}
\subfigure[Iberia]{\includegraphics[width=0.25\textwidth]{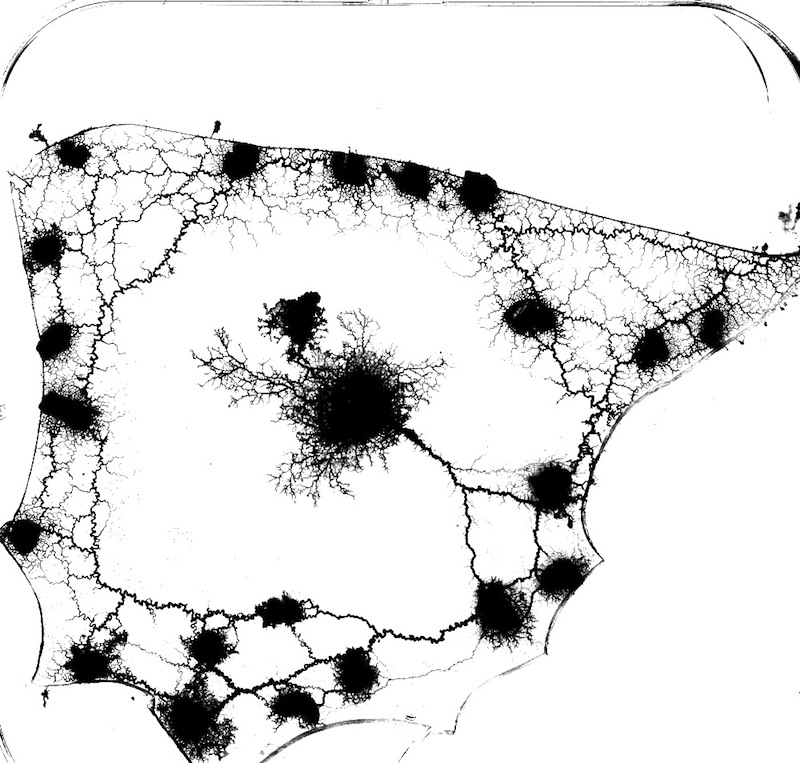}}
\subfigure[Malaysia]{\includegraphics[width=0.25\textwidth]{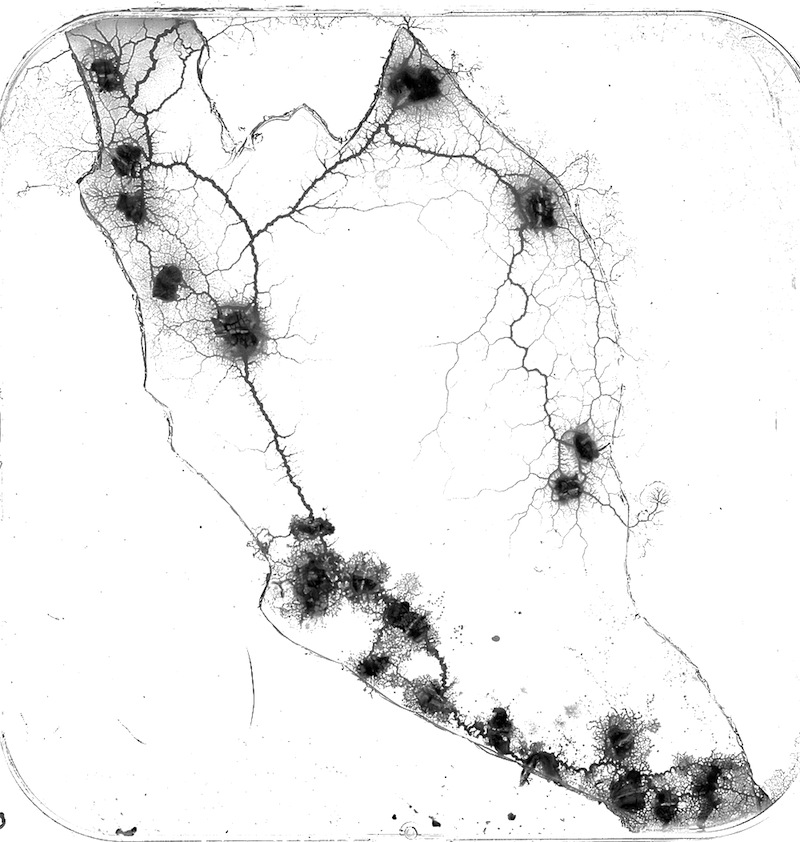}}
\subfigure[Mexico]{\includegraphics[width=0.25\textwidth]{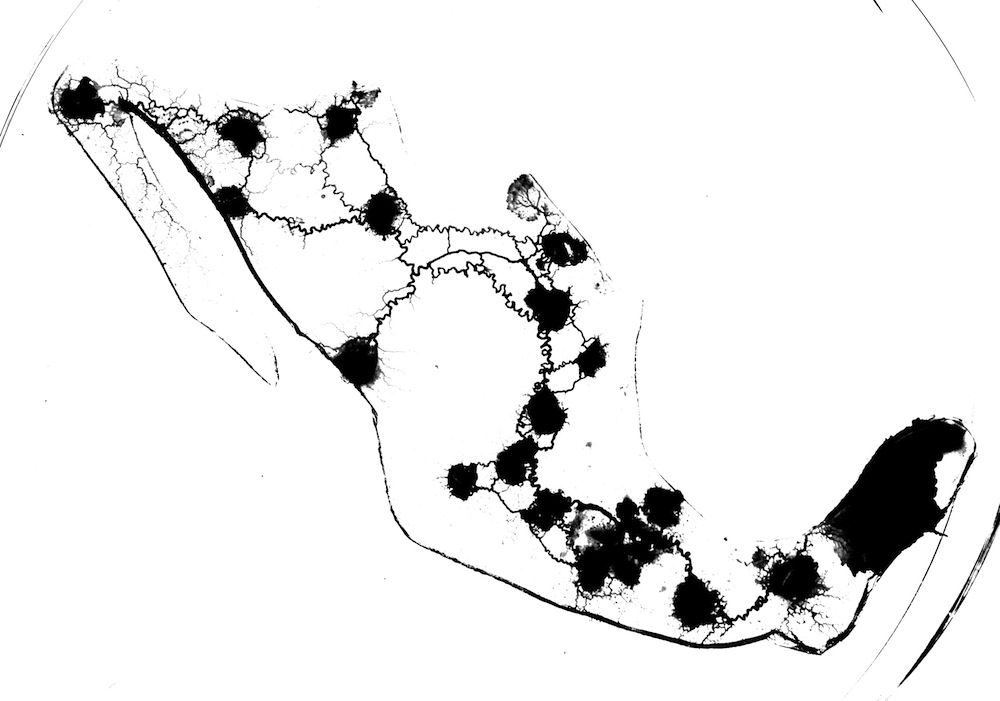}} \hspace{0.1cm}
\subfigure[The Netherlands]{\includegraphics[width=0.25\textwidth]{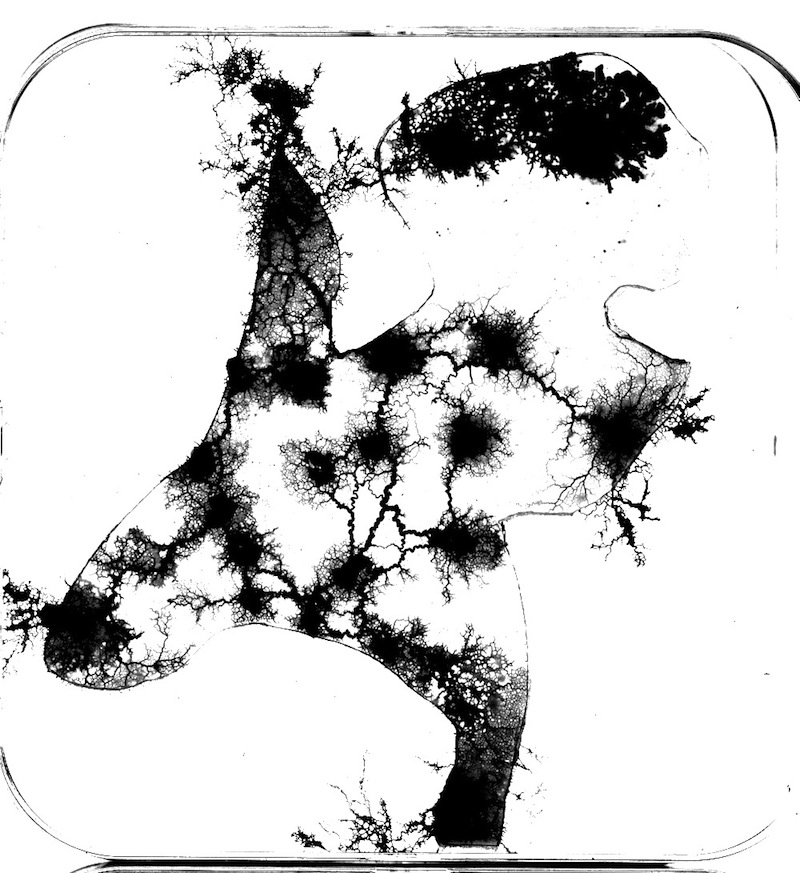}}
\subfigure[USA]{\includegraphics[width=0.25\textwidth]{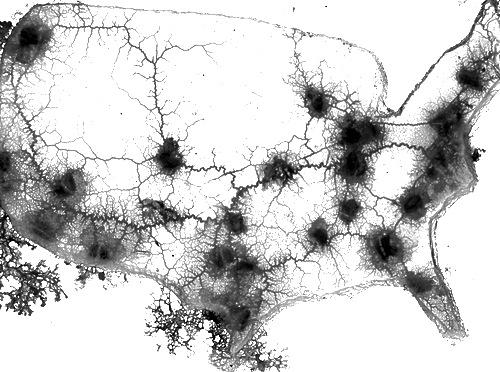}}
\caption{Exemplar configurations of protoplasmic networks developed by slime mould \emph{P. polycephalum} on 
major urban areas $\mathbf{U}$ obtained in experimental laboratory studies.}
\label{physarumexamples}
\end{figure}

In laboratory experiments~\cite{PhysarumIberia}--\cite{PhysarumChina},\cite{PhysarumItaly} we considered fourteen regions:
Australia,  Africa, Belgium, Brazil, Canada, China, Germany, Iberia, Italy, Malaysia, Mexico, The Netherlands, UK, and USA. 
Agar plates, 
2\% agar gel (Select agar, Sigma Aldrich), were cut in a shape of any particular region and placed in polyestyrene square Petri 
dishes $120 \times 120$~mm or $220 \times 220$~mm. For each region we choose the most populated urban areas $\mathbf U$ 
scaled down locations of which were projected onto agar gel. Numbers of the areas selected for each country are as follows: 
Africa 	$n=35$, 
Australia	$n=25$,
Belgium	$n=21$,
Brazil	$n=21$,
Canada	$n=16$,
China	$n=31$,
Germany	$n=21$,
Iberia	$n=23$,
Italy $n=11$,
Malaysia	$n=20$,
Mexico	$n=19$,
Netherlands	$n=21$,
UK	$n=10$,
USA	$n=20$, see detailed configurations in~\cite{PhysarumIberia}--\cite{PhysarumChina},\cite{PhysarumItaly}.

 At the beginning of each experiment a piece of plasmodium, usually already 
attached to an oat flake, is placed in the capital city. The Petri dishes with plasmodium were kept in darkness, at 
temperature 22-25~C$^{\text o}$, except for observation and image recording (Fig.~\ref{onmaps}).  Periodically (usually in 12~h or 24~h 
intervals) the dishes were scanned in Epson Perfection 4490. Examples of typical protoplasmic networks recorded in laboratory experiments are shown in Fig.~\ref{physarumexamples}.  Detailed examples and scenarios of colonisation of various regions by the slime mould are provided in original 
papers~\cite{PhysarumIberia}--\cite{PhysarumChina},\cite{PhysarumItaly}.

\begin{figure}[!tbp]
\centering
\subfigure[]{\includegraphics[width=0.49\textwidth]{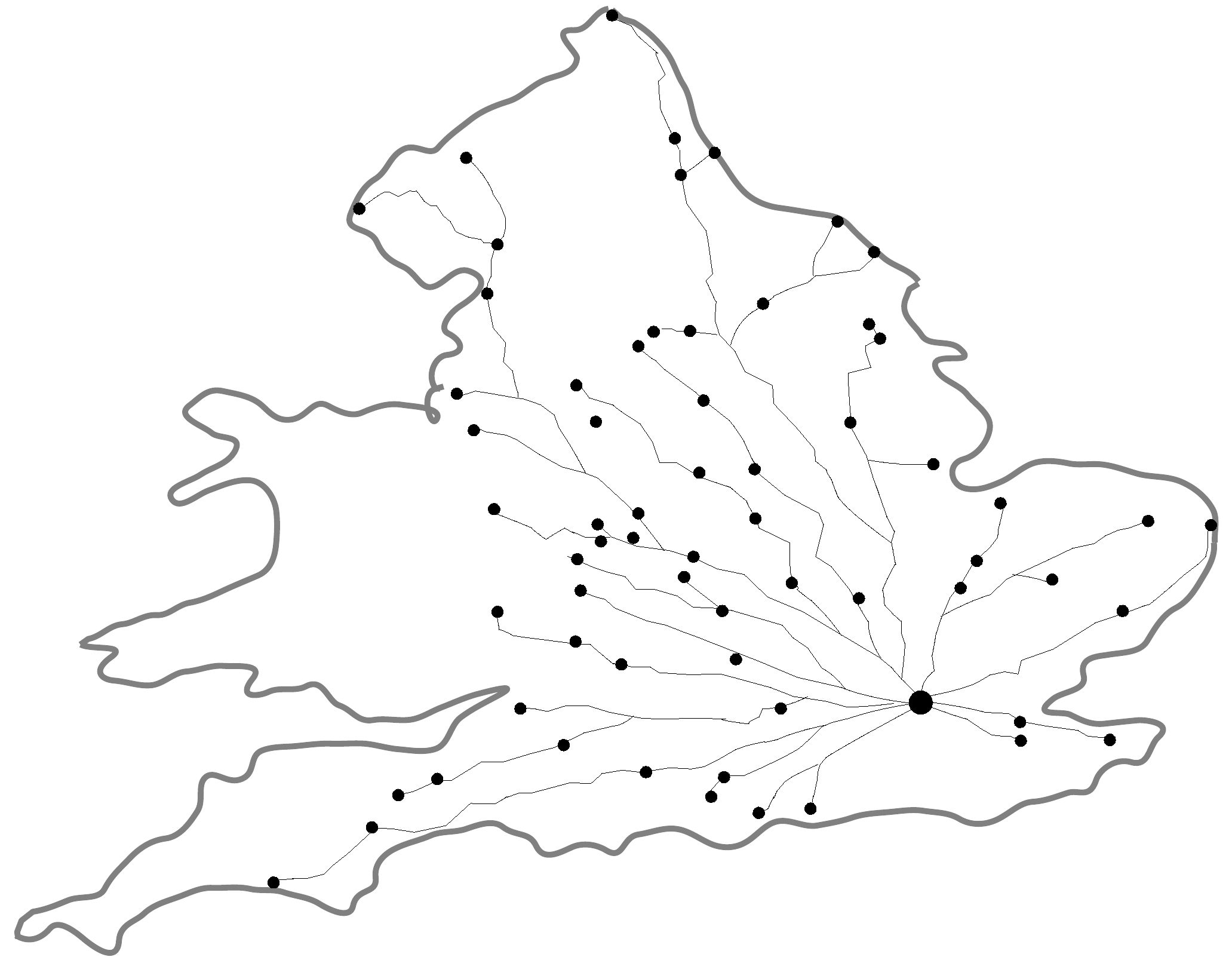}\label{turnpikes1}}
\subfigure[]{\includegraphics[width=0.35\textwidth]{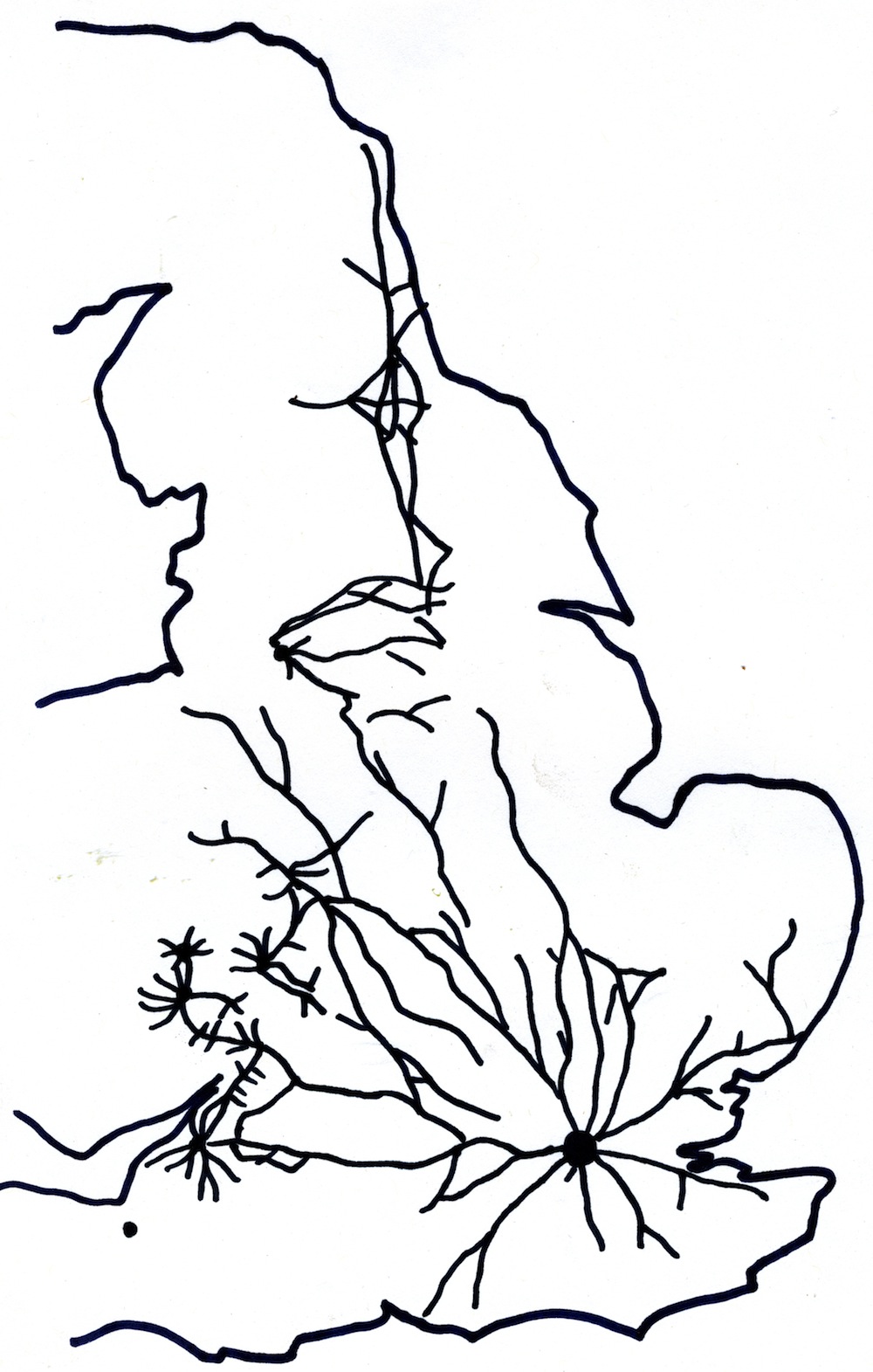}\label{turnpikes2}}
\caption{Turnpike road network. (a)~A scheme of  London turnpike network in England, c. 1720. Modified from~\cite{bogart_2007}.
(b)~A scheme of the England turnpike network in 1750. Modified from~\cite{davies}.}
\label{turnpikes}
\end{figure}

In our experiments we always inoculated plasmodium of \emph{P. polycephalum} in a capital of any particular country. This is because
in the majority of cases --- but not in all cases indeed --- capitals are the most populated and industrially developed urban areas, and the road 'diffusion' in ancient times was usually originated from the capital city. For example, see a rough scheme of  turnpikes in England (Fig.~\ref{turnpikes1}), which demonstrates a classical growth pattern, typical for fungi, myxomecetes and bacterial colonies.  Moreover, we can even see evidence of secondary growth
from other cities, e.g. in Fig.~\ref{turnpikes2} we see the main turnpike network growing from London and several sub-networks growing 
from Bristol, Ross, Leominster, Worcester, and Manchester. 

Unlike bacterial colonies and fungal mycelia, the Physarum plasmodium is 
able to shift the mass of its body plan by the adaptive assembly and 
disassembly of its protoplasmic tube network and utilise the transport of 
structural components within the network to the leading edge of growth. The 
morphology of the plasmodium is thus less dependent on the initial 
inoculation site and the active zone of growth can move throughout the 
environment as the plasmodium forages for food.

Also, from our previous experimental studies we know that when plasmodium is inoculated in 
every point of a given planar set, the protoplasmic network formed approximates the Delaunay triangulation of the set~\cite{adamatzky_ppl_2008,shirakawa_2009}. Neither of the motorway networks considered match the  Delaunay triangulation of major urban areas, thus simultaneous  inoculation in all urban areas would not bring any additional benefits.

\begin{figure}[!tbp]
\centering
\subfigure[Africa]{\includegraphics[scale=0.1]{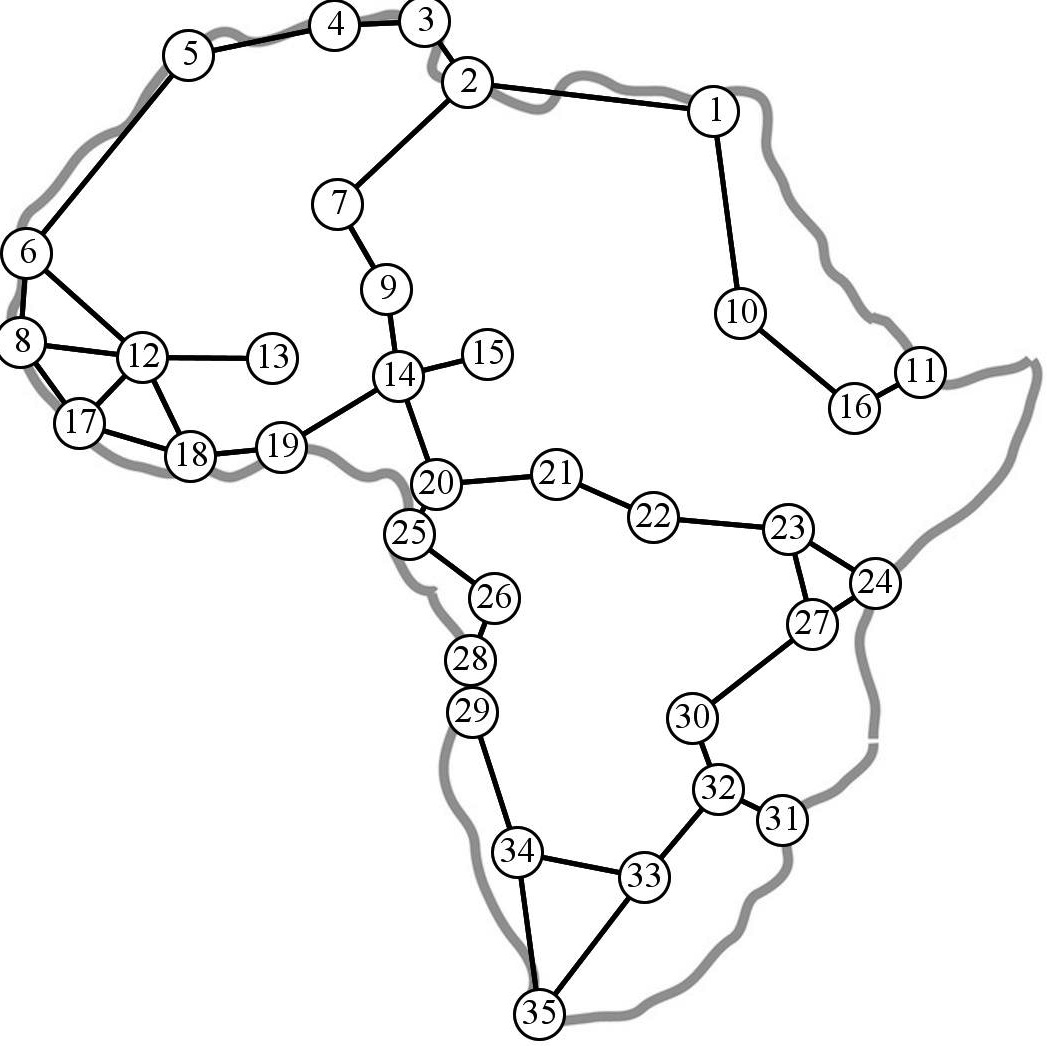}\label{AfricaP}}
\subfigure[Australia]{\includegraphics[scale=0.1]{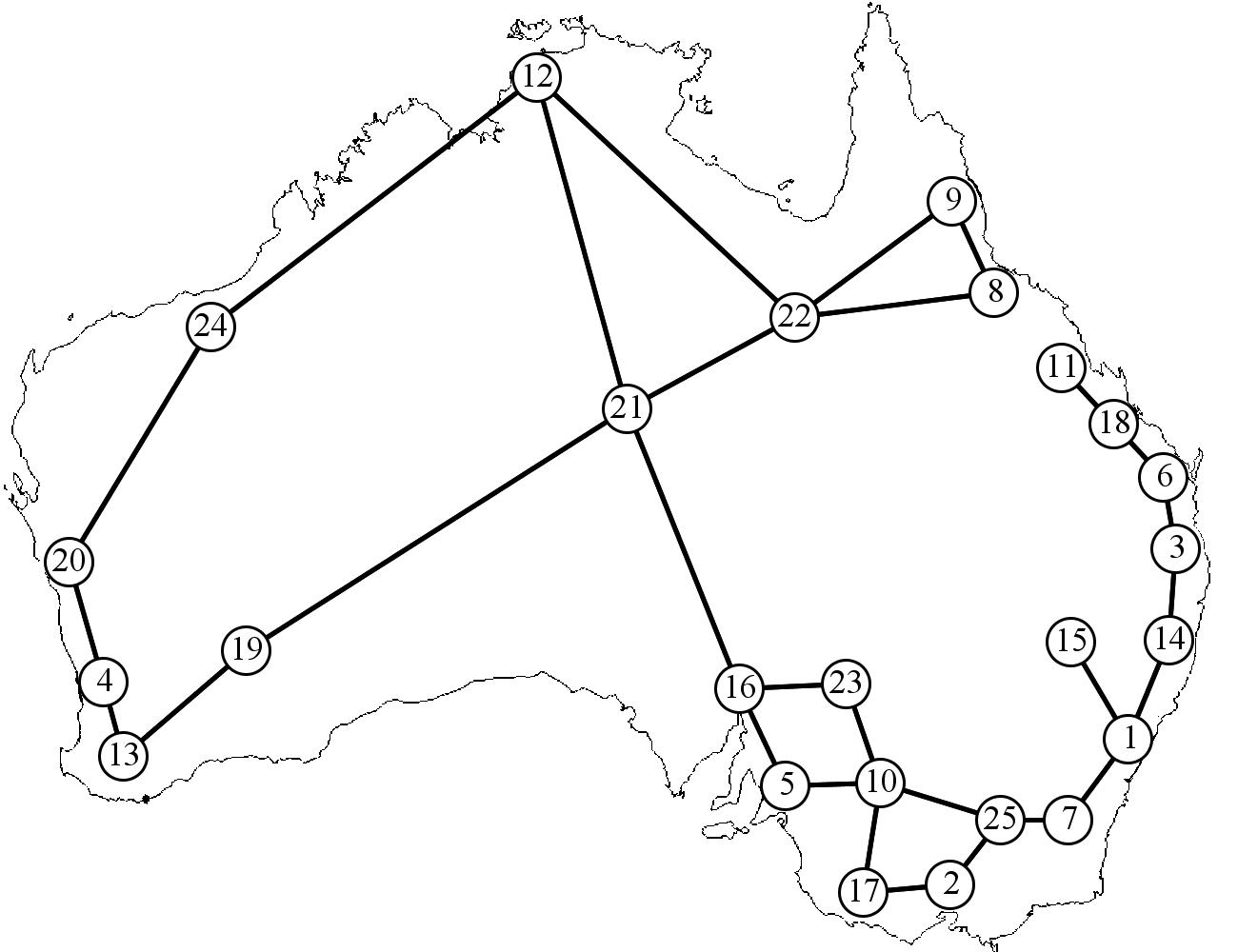}\label{AustraliaP}}
\subfigure[Belgium]{\includegraphics[scale=0.1]{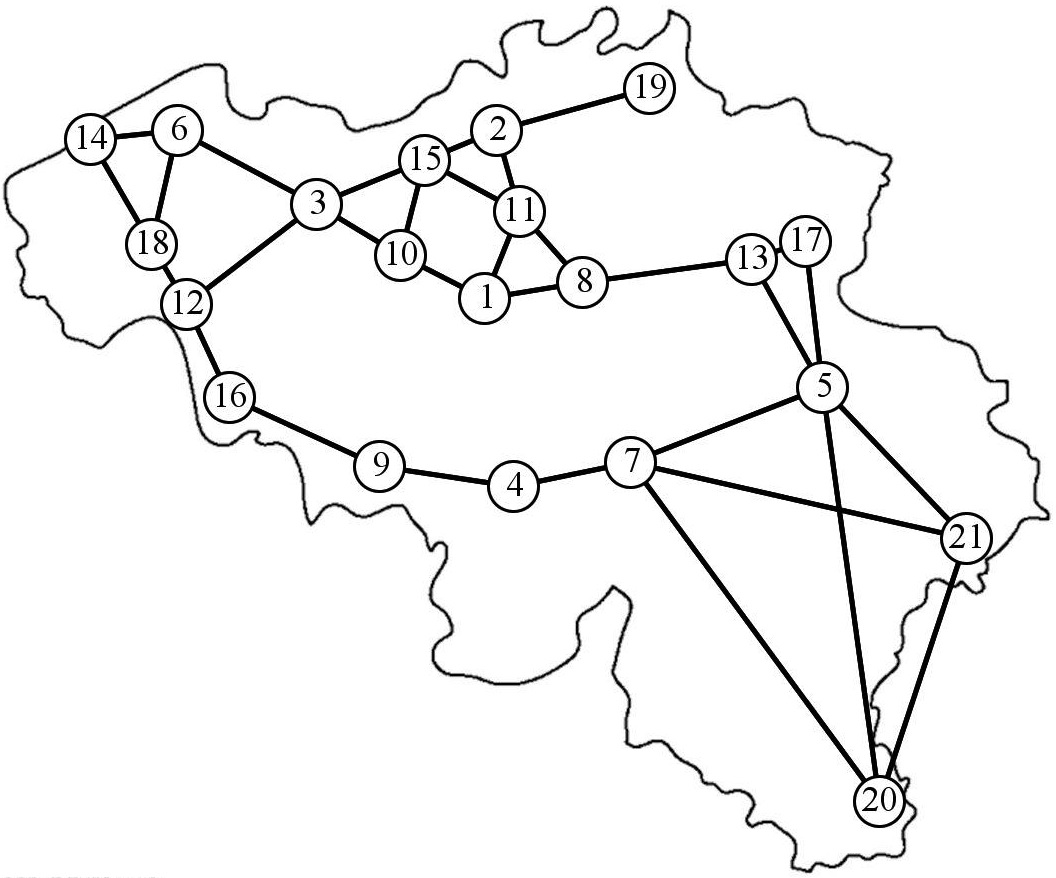}\label{BelgiumP}}
\subfigure[Brazil]{\includegraphics[scale=0.1]{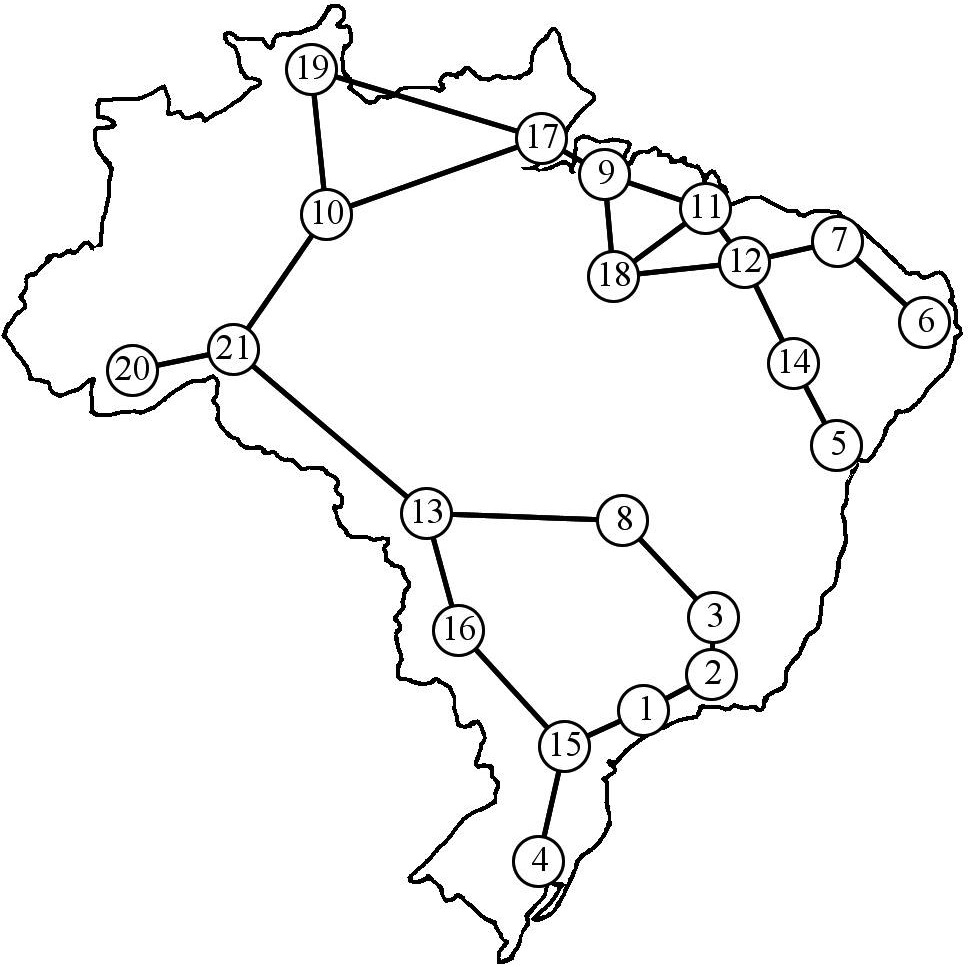}\label{BrazilP}}
\subfigure[Canada]{\includegraphics[scale=0.1]{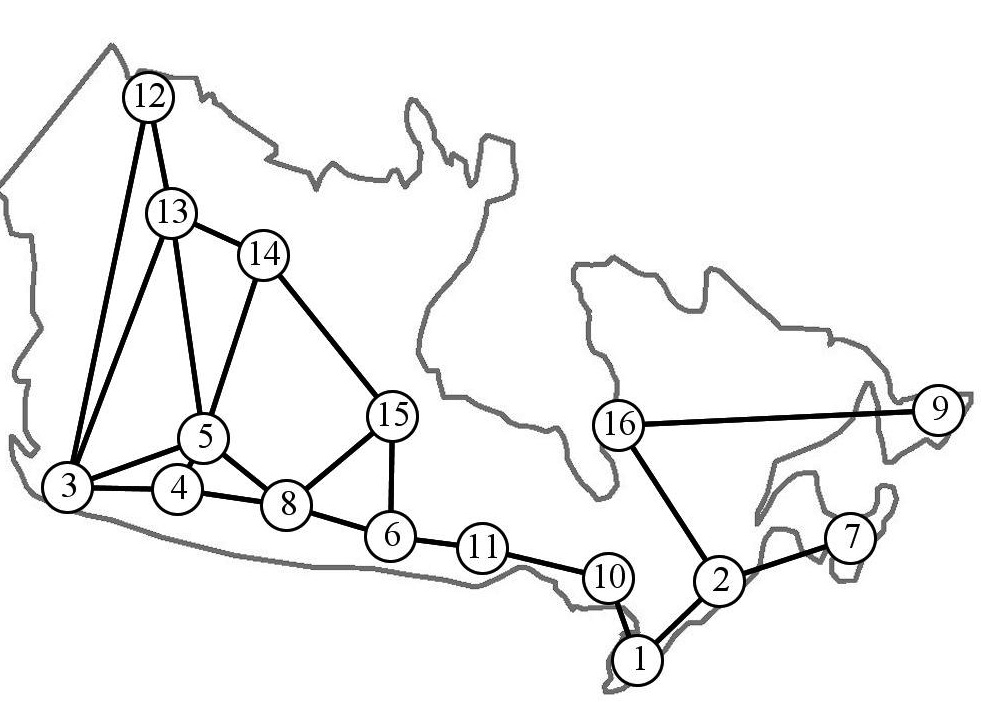}\label{CanadaP}}
\subfigure[China]{\includegraphics[scale=0.1]{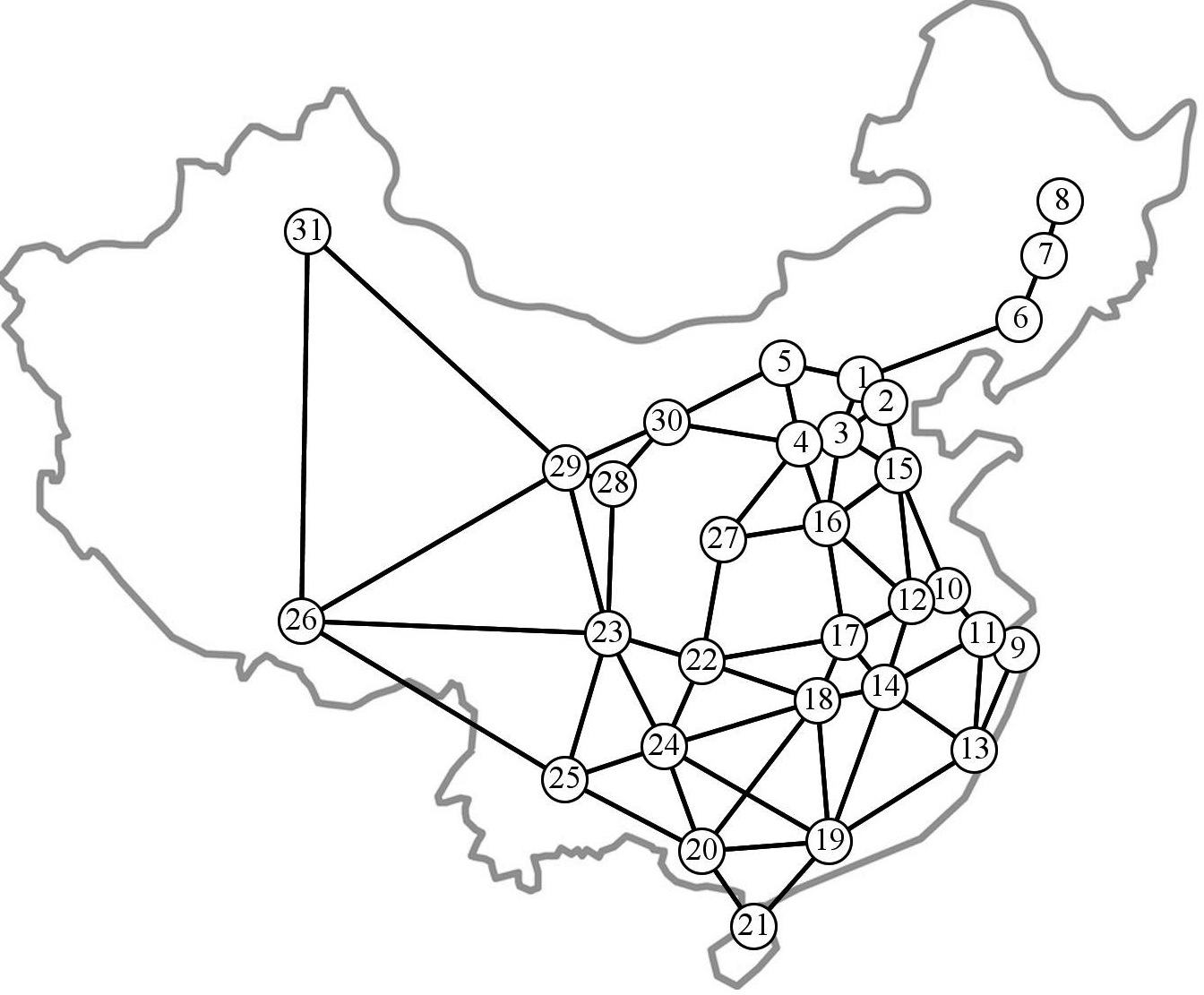}\label{ChinaP}}
\subfigure[Germany]{\includegraphics[scale=0.1]{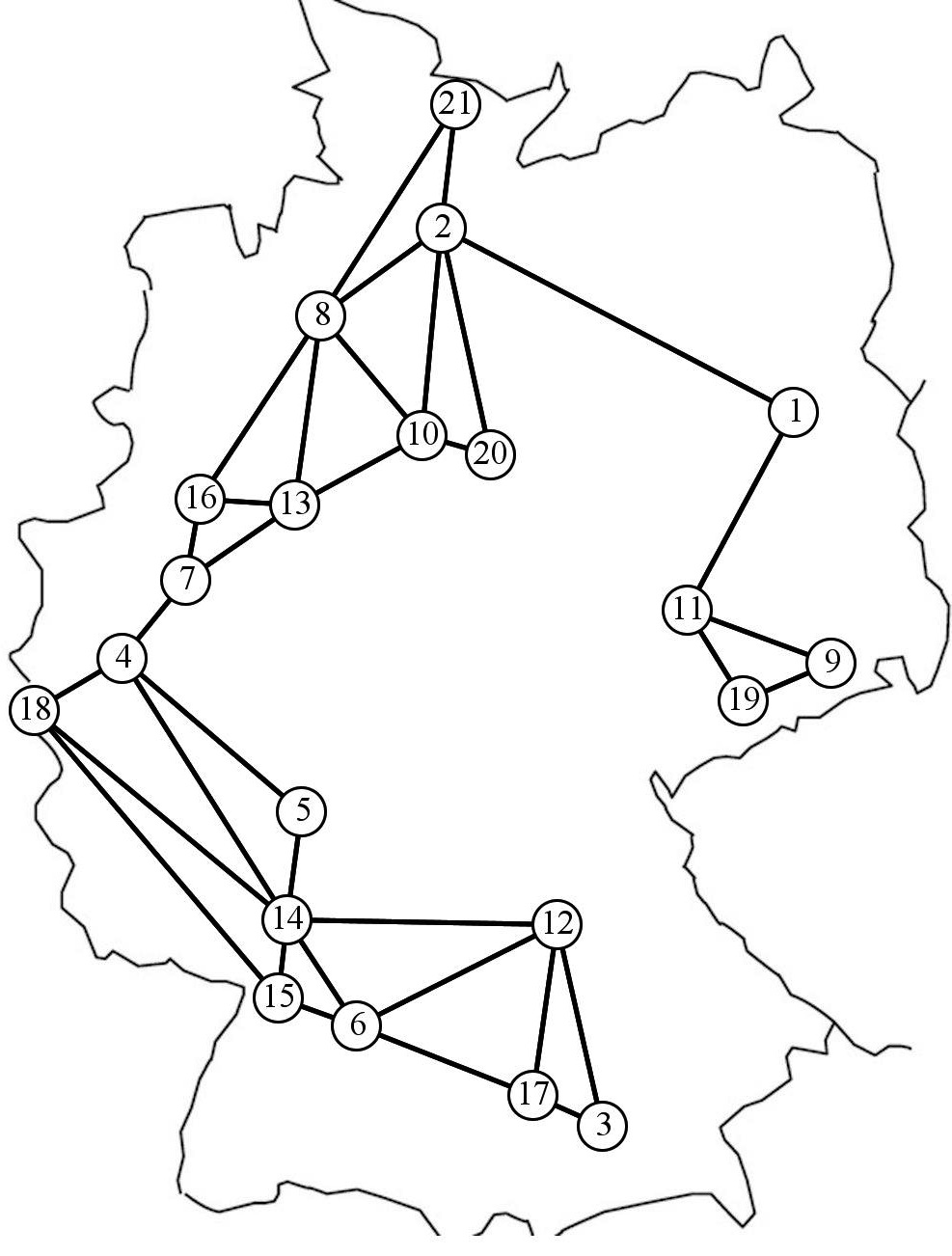}\label{GermanyP}}
\subfigure[Iberia]{\includegraphics[scale=0.1]{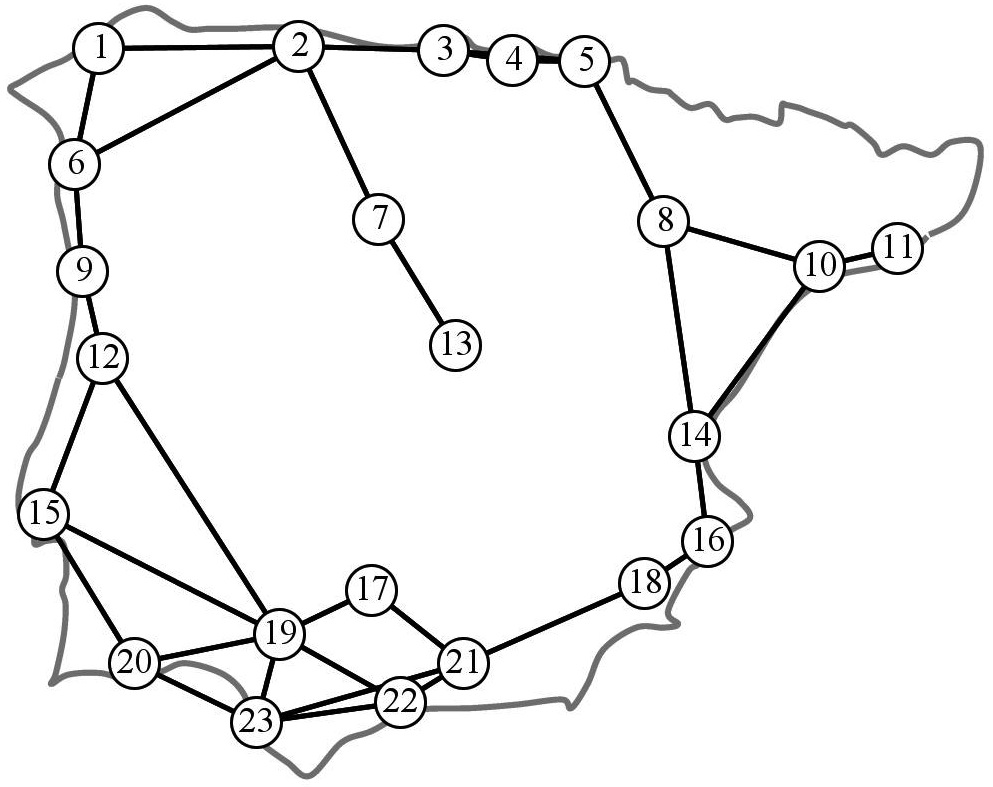}\label{IberaiP}}
\subfigure[Italy]{\includegraphics[scale=0.1]{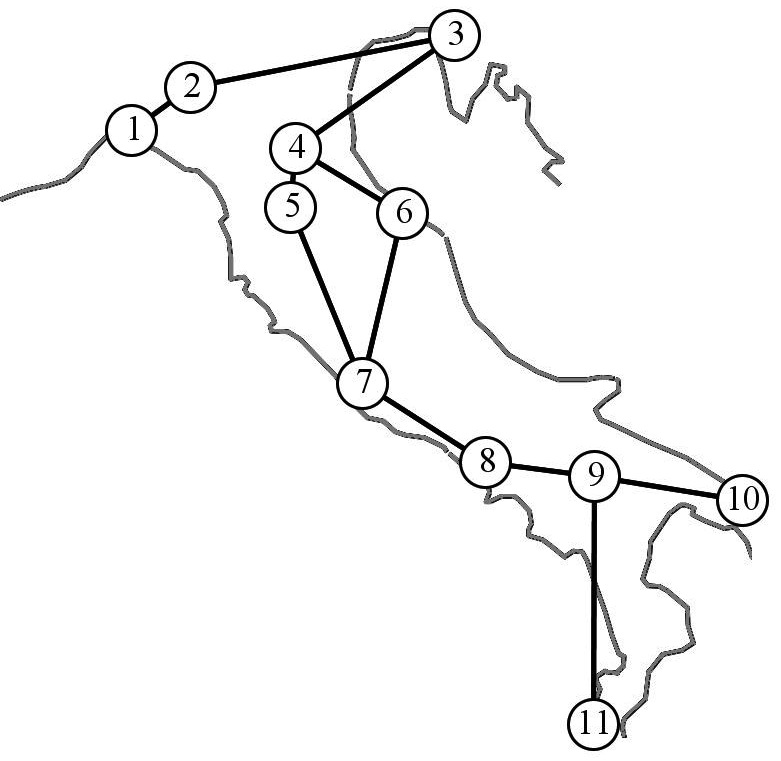}\label{ItalyP}}
\subfigure[Malaysia]{\includegraphics[scale=0.1]{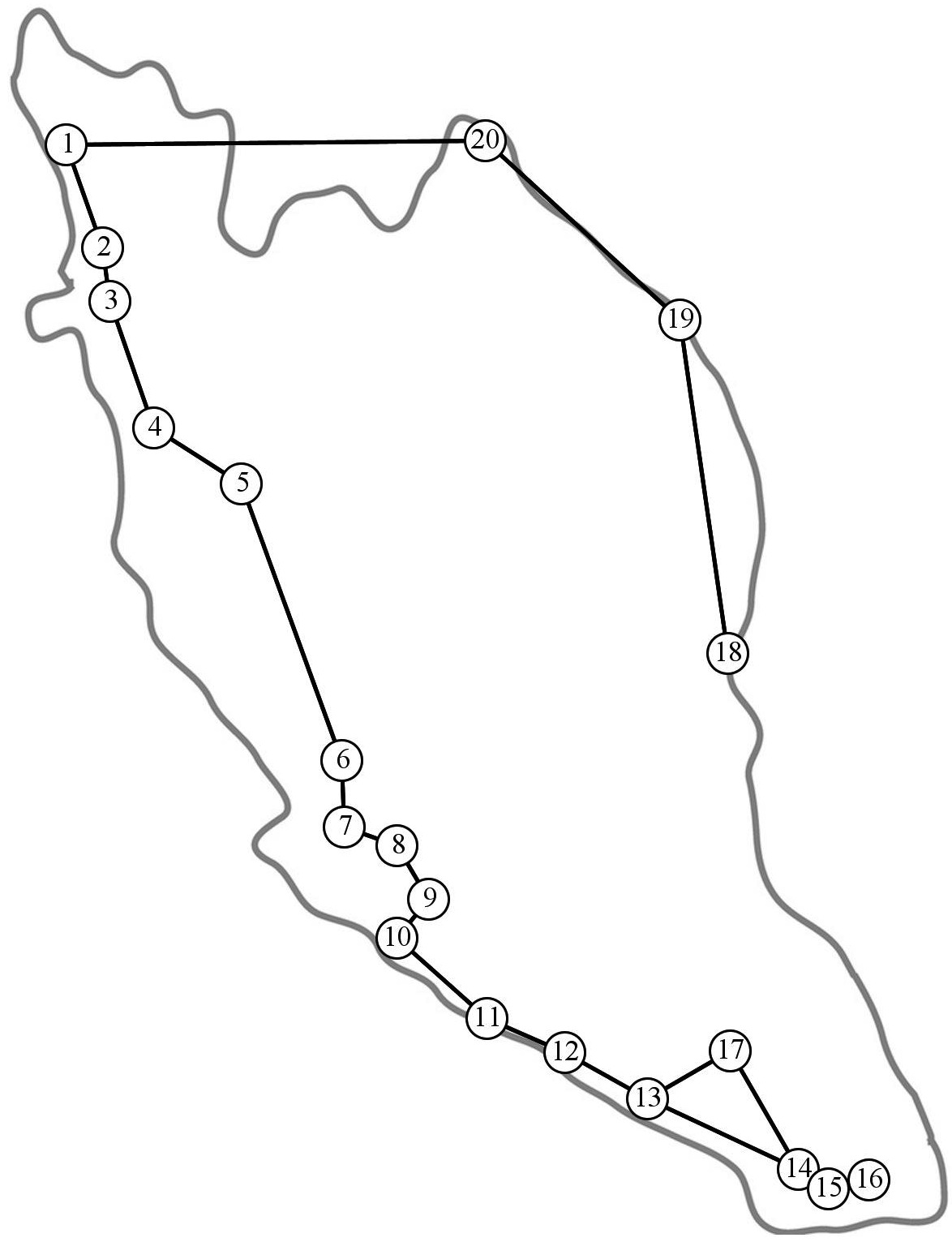}\label{MalaysiaP}}
\subfigure[Mexico]{\includegraphics[scale=0.1]{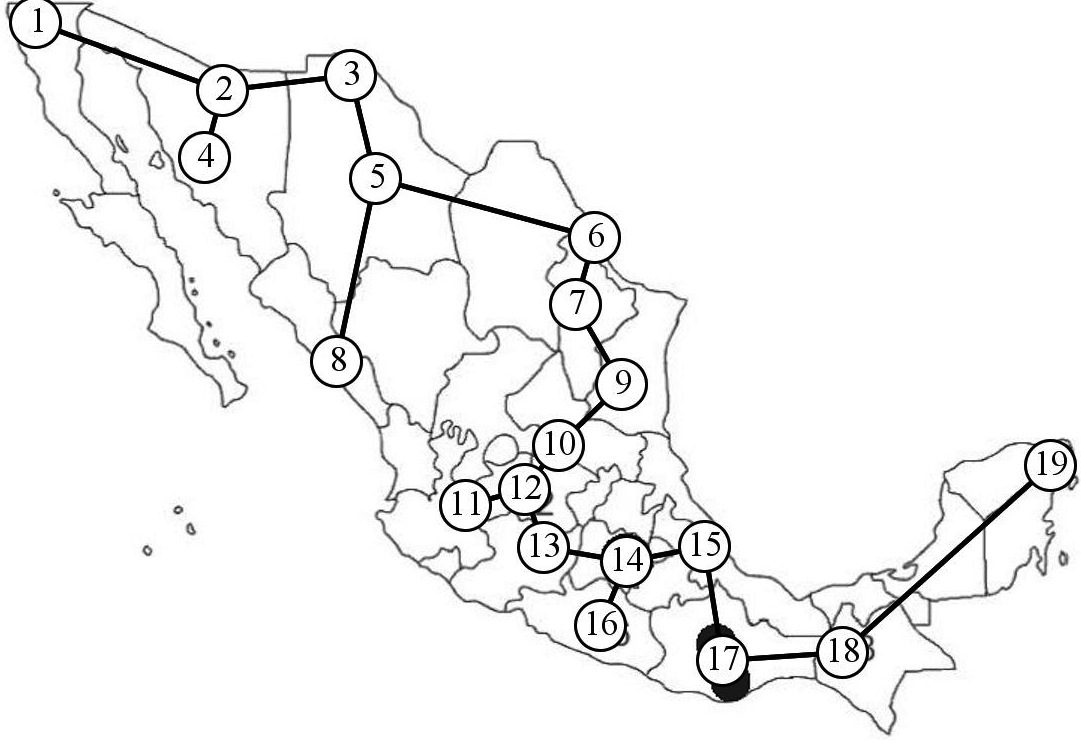}\label{MexicoP}}
\subfigure[The Netherlands]{\includegraphics[scale=0.1]{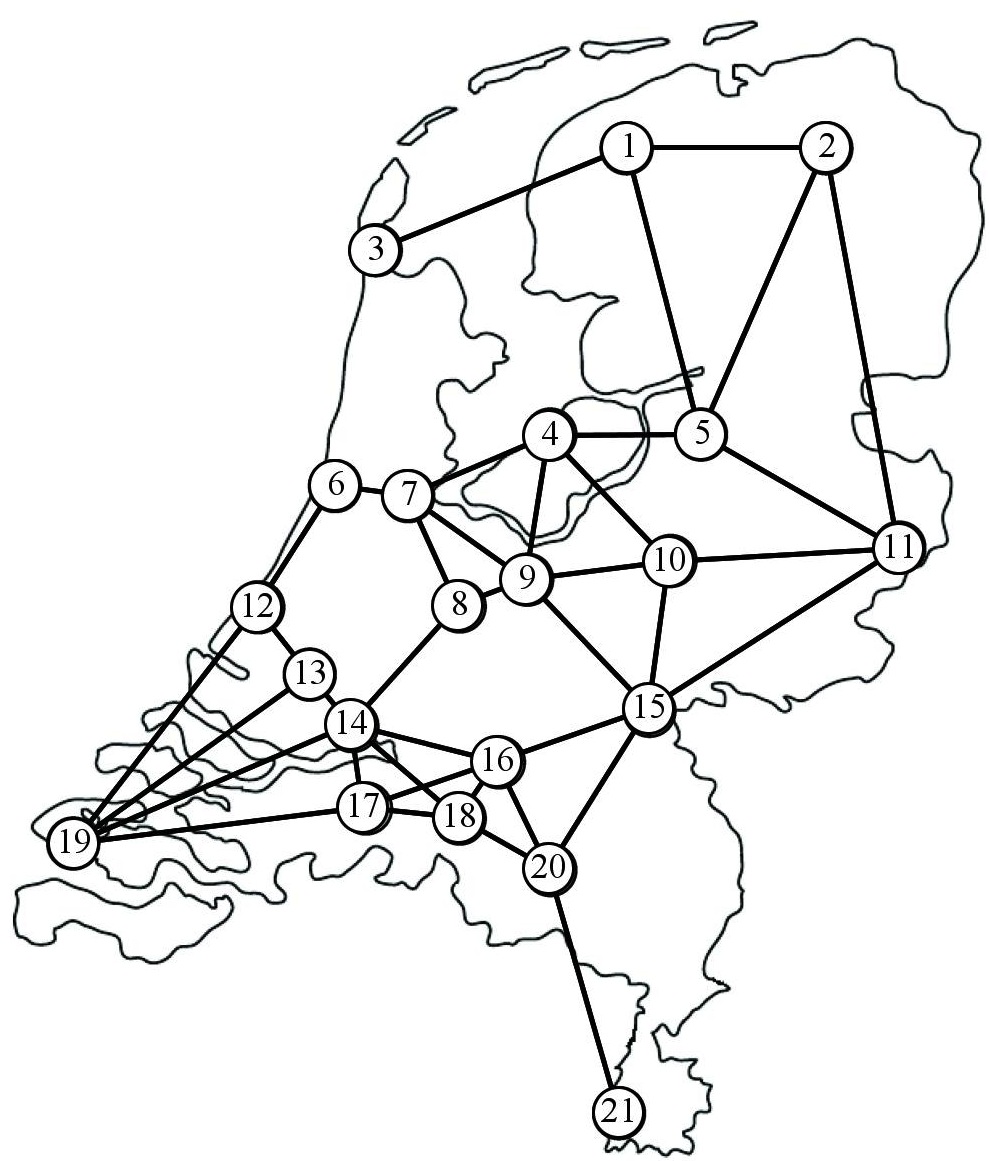}\label{NetherlandsP}}
\subfigure[UK]{\includegraphics[scale=0.1]{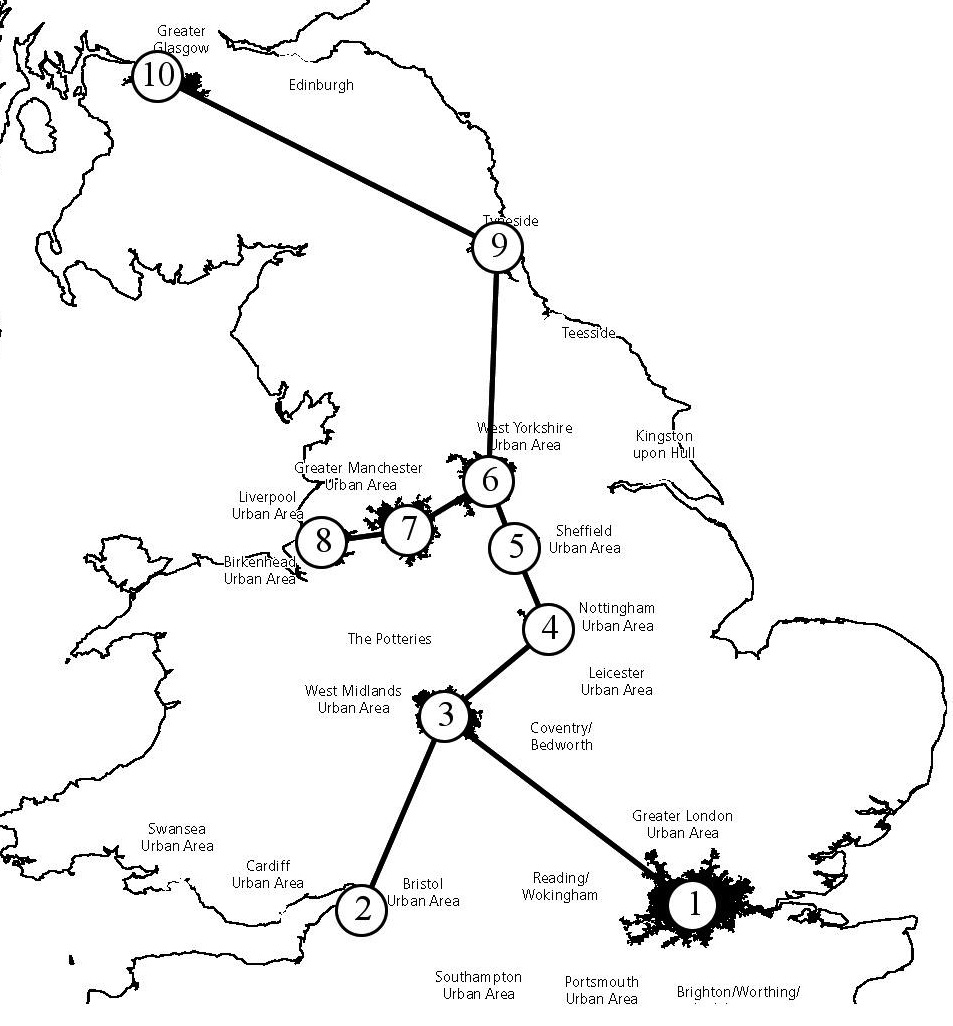}\label{UKP}}
\subfigure[USA]{\includegraphics[scale=0.1]{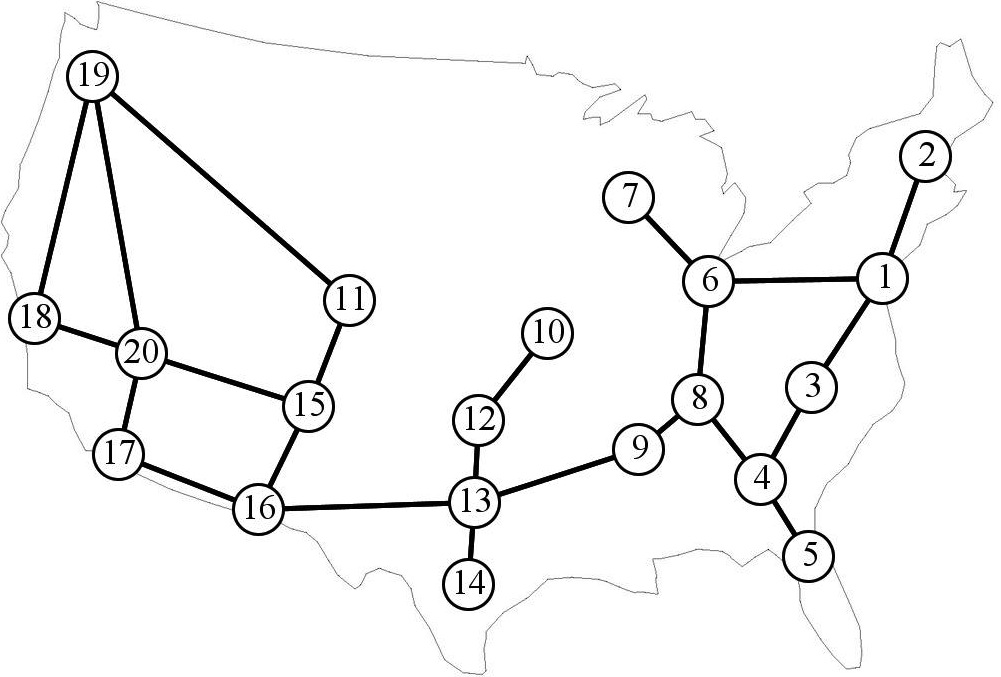}\label{USAP}}
\caption{Physarum graphs $\mathbf{P}(\theta')$ for highest values of $\theta$ which do not make the graphs disconnected. }
\label{physarumgraphs}
\end{figure}

As with every living creature, the plasmodium of \emph{P. polycephalum} does not always repeat 
its foraging pattern. To generalise our experimental results we constructed a Physarum graph with weighted-edges.
A Physarum graph  is a tuple ${\mathbf P} = \langle {\mathbf U}, {\mathbf E}, w  \rangle$, where
$\mathbf U$ is a set of  urban areas, $\mathbf E$ is a set edges, and $w: {\mathbf E} \rightarrow [0,1]$ associates each edge of $\mathbf{E}$ with  a frequency (or weight) of the edge occurrence in laboratory experiments. For every two regions $a$ and $b$ from $\mathbf U$ there is an edge connecting $a$ and $b$ if a plasmodium's protoplasmic link is recorded at least in one of $k$ experiments, and the edge $(a,b)$ has a weight calculated as a ratio of experiments where protoplasmic link $(a,b)$ occurred in the total number of experiments $k$.  We do not take into account the exact configuration of the protoplasmic tubes but merely their existence. In original papers~\cite{PhysarumIberia}--\cite{PhysarumChina},\cite{PhysarumItaly} we dealt with threshold Physarum graphs $\mathbf{P}(\theta)  = \langle  {\mathbf U}, T({\mathbf E}), w, \theta \rangle$. The threshold Physarum graph is obtained from the Physarum graph by the transformation: $T({\mathbf E})=\{ e \in {\mathbf E}: w(e) > \theta \}$. That is, all edges with weights less than or equal to $\theta$ are removed.  With the increase of $\theta$ in a family of threshold Physarum graphs $\{{\mathbf P}(\theta), \theta=0,1,2,\cdots, k-1 \}$ the graphs undergo the following transitions (see country-specific details in~\cite{PhysarumIberia}--\cite{PhysarumChina},\cite{PhysarumItaly}): 
\centerline{non-planar connected $\rightarrow$ planar connected $\rightarrow$ disconnected  $\rightarrow$ all nodes are isolated}.

\begin{figure}[!tbp]
\centering
\subfigure[$\theta'$ vs number of nodes]{\includegraphics[width=0.49\textwidth]{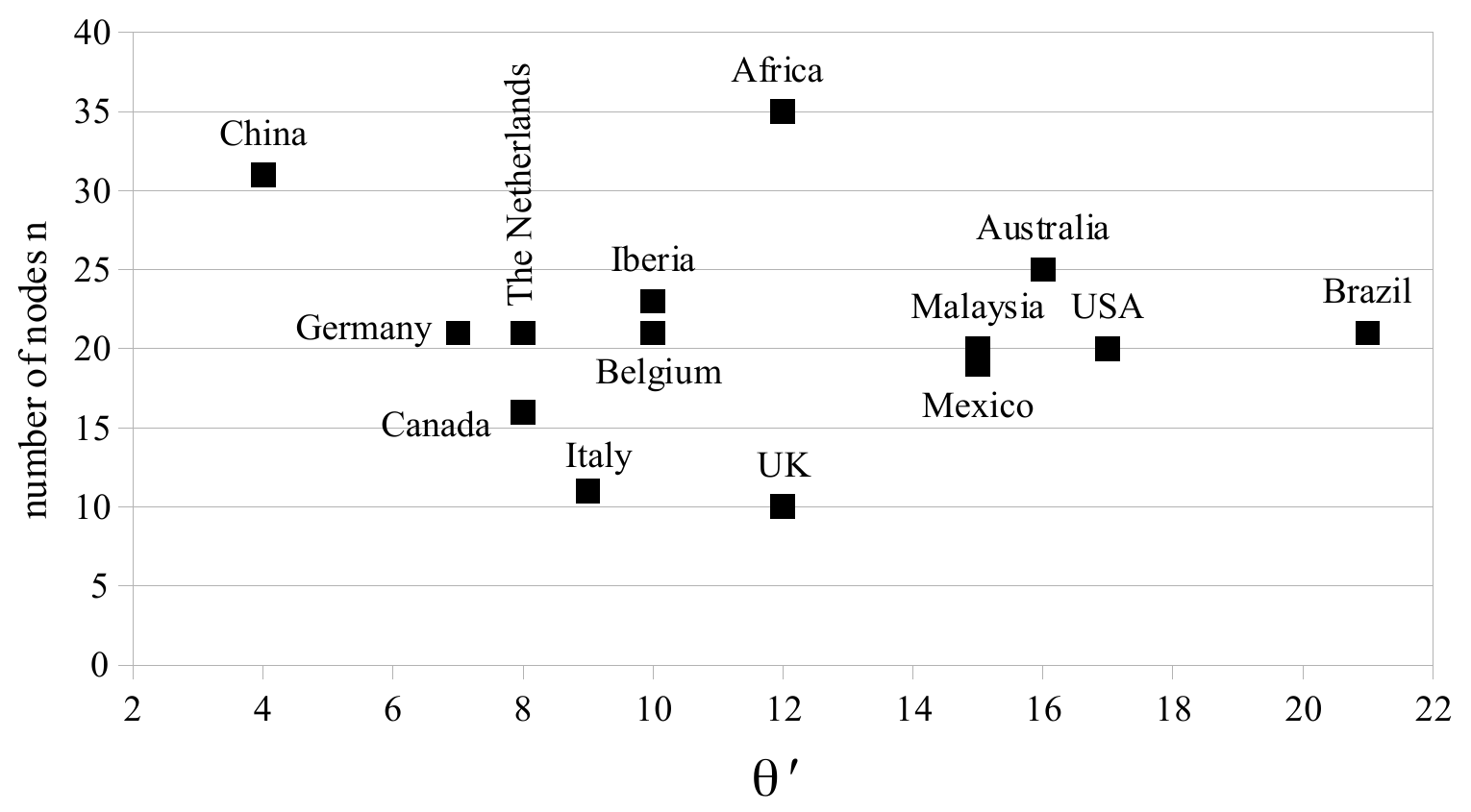}\label{ThetaVsNodes}}
\subfigure[Matching vs economy]{\includegraphics[width=0.49\textwidth]{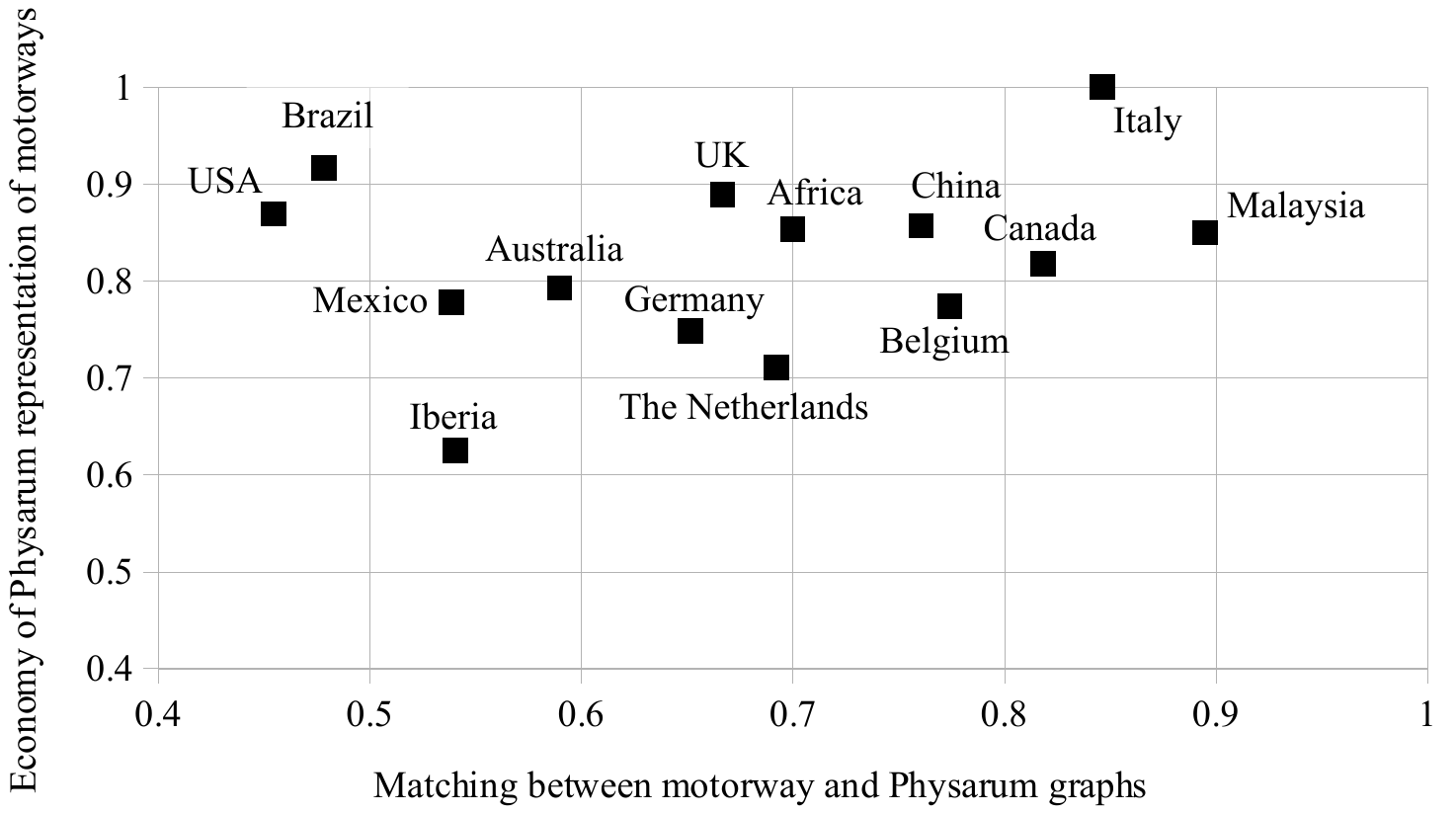}\label{MatchingEconomy}}
\caption{
(a)~$\theta'$ versus number of nodes.
(d)~Matching $\mu$ versus economy $\epsilon$.
}
\label{theteANDmatching}
\end{figure}

In the present paper we consider only 'stressed' Physarum graphs $\mathbf{P}(\theta')$, which have maximum possible values
$\theta'$ and yet remain connected: $\theta'= \max\{\theta: \mathbf{P}(\theta)=\text{connected}\}$. 
These Physarum graphs are shown in Fig.~\ref{physarumgraphs}. Values of $\theta'$ for studied regions are illustrated in Fig.~\ref{ThetaVsNodes}. 

\begin{figure}[!tbp]
\centering
\subfigure[Africa]{\includegraphics[scale=0.1]{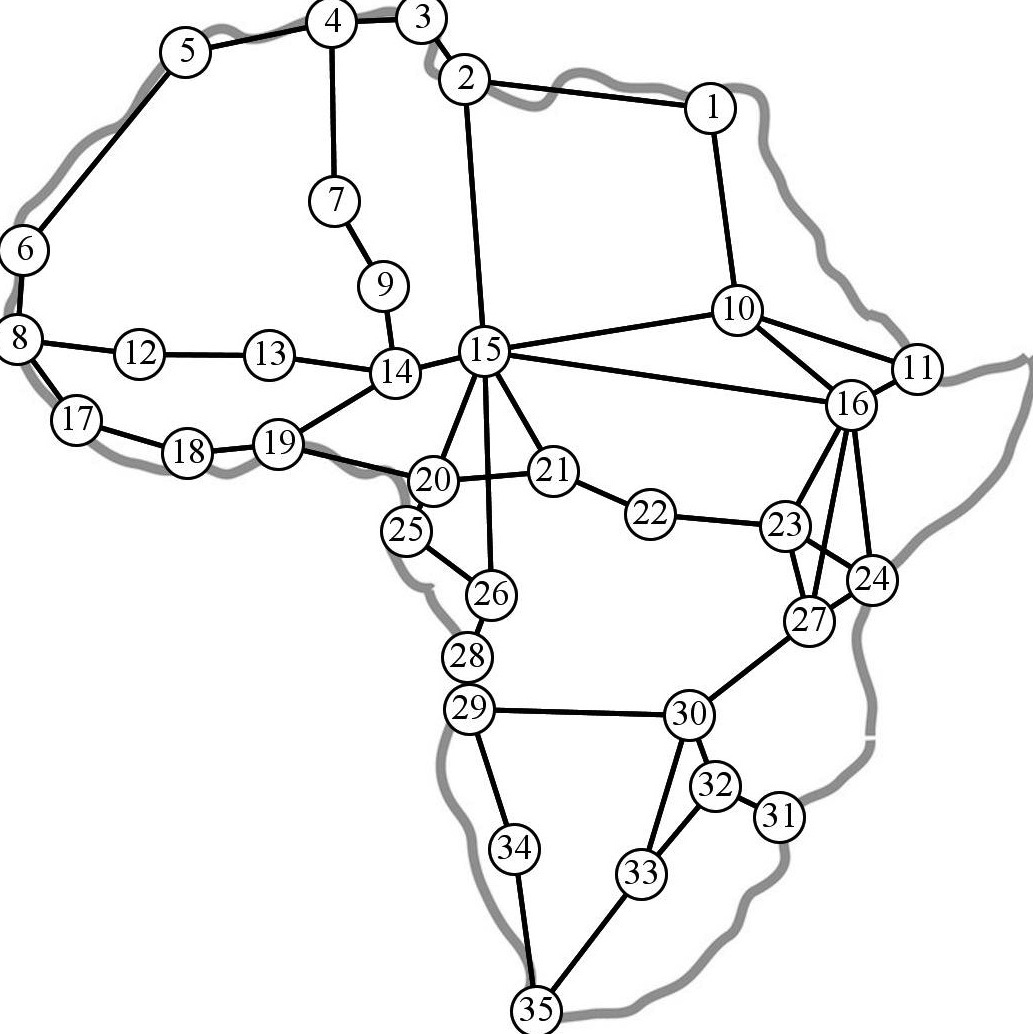}\label{AfricaH}}
\subfigure[Australia]{\includegraphics[scale=0.1]{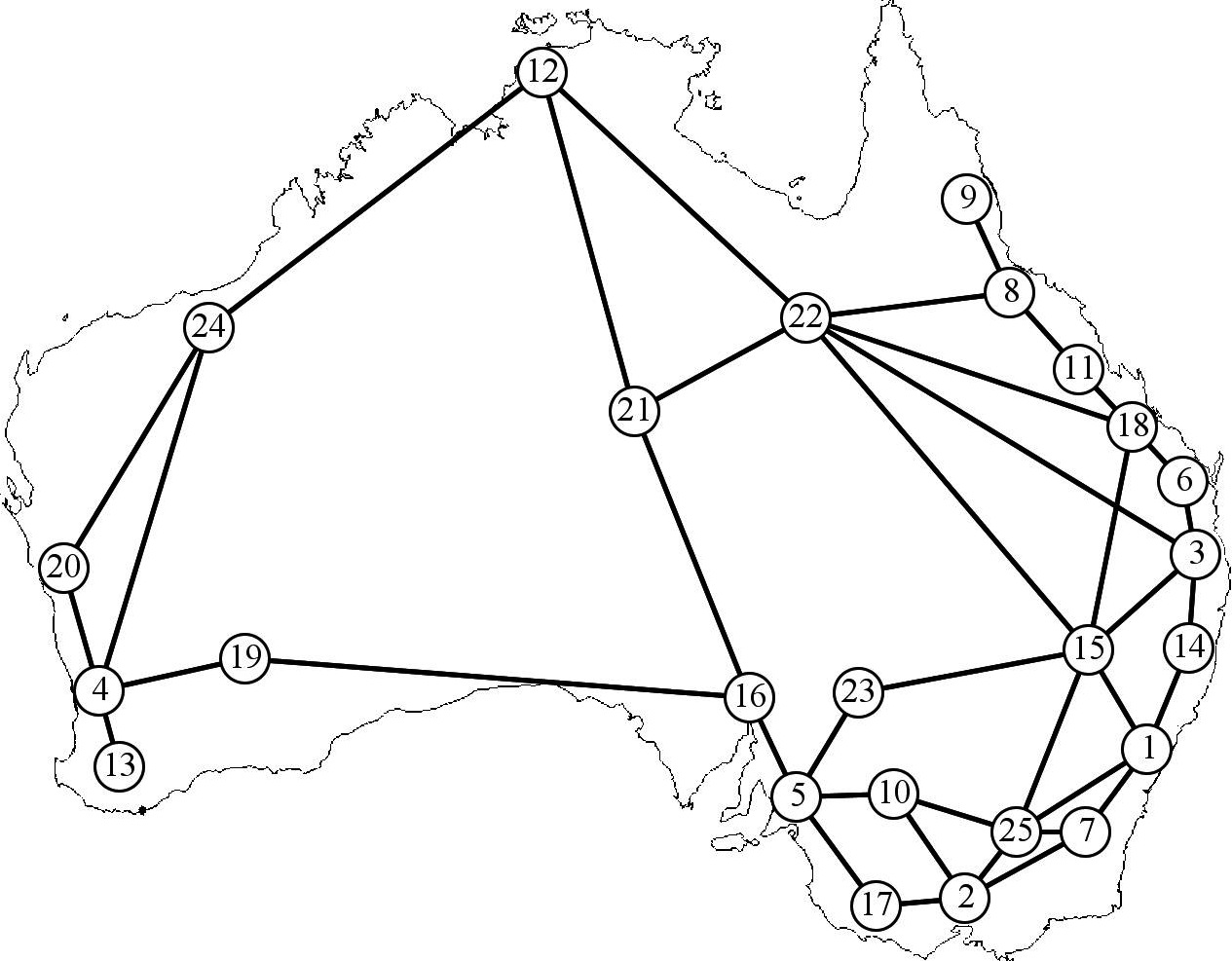}\label{AustraliaH}}
\subfigure[Belgium]{\includegraphics[scale=0.1]{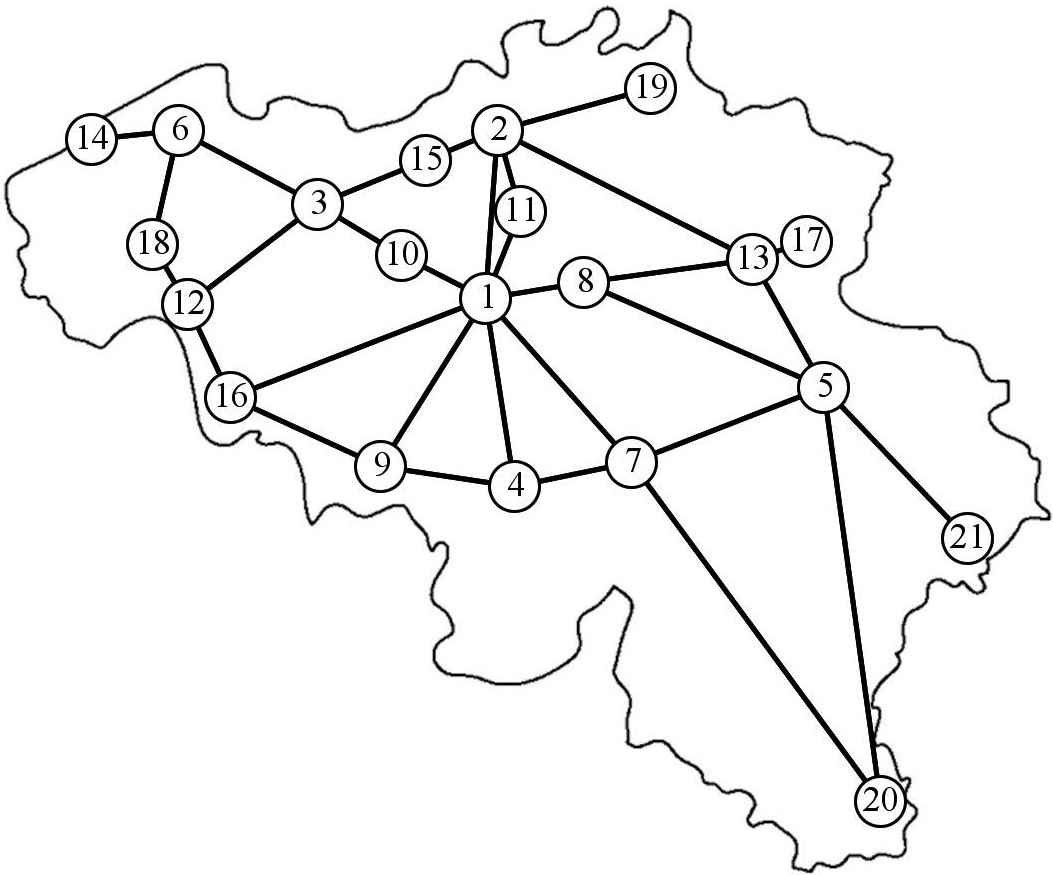}\label{BelgiumH}}
\subfigure[Brazil]{\includegraphics[scale=0.1]{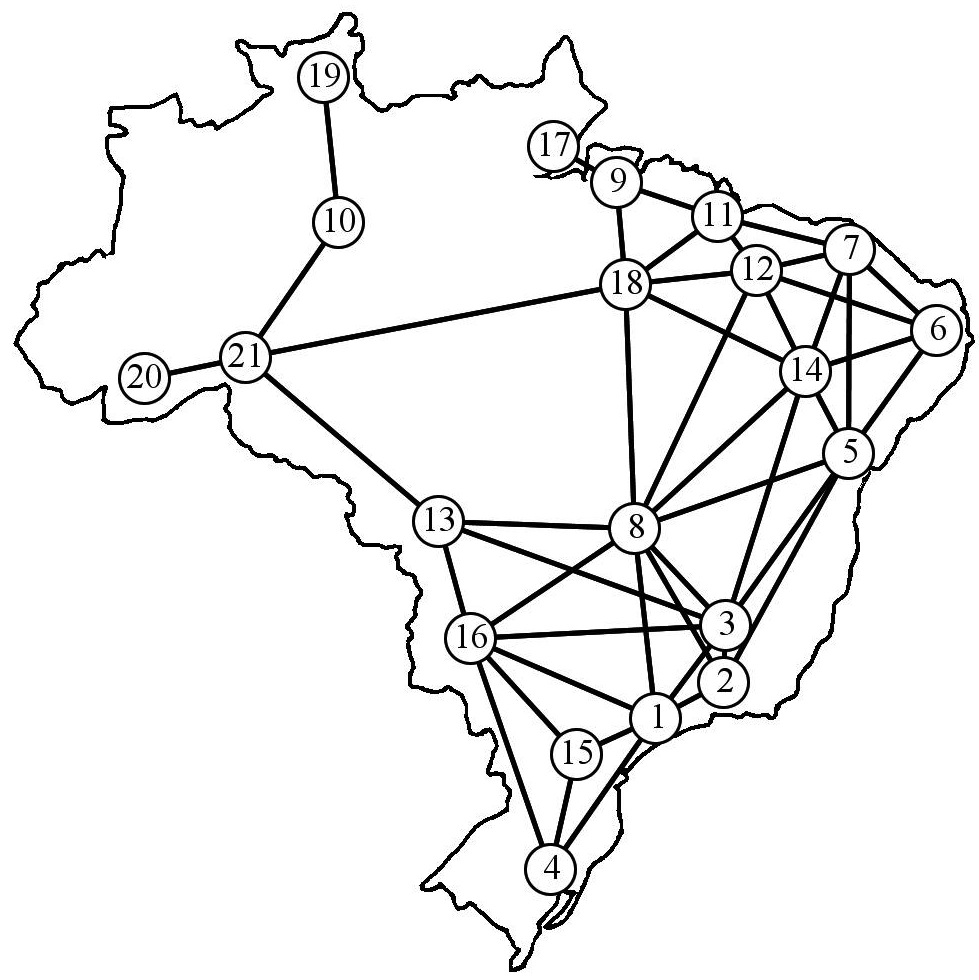}\label{BrazilH}}
\subfigure[Canada]{\includegraphics[scale=0.1]{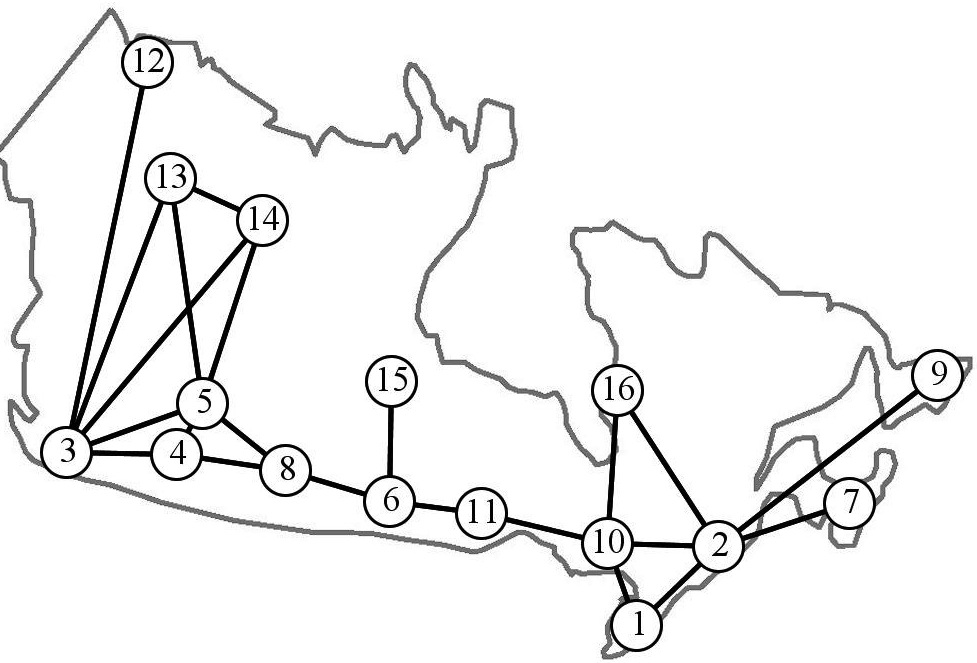}\label{CanadaH}}
\subfigure[China]{\includegraphics[scale=0.1]{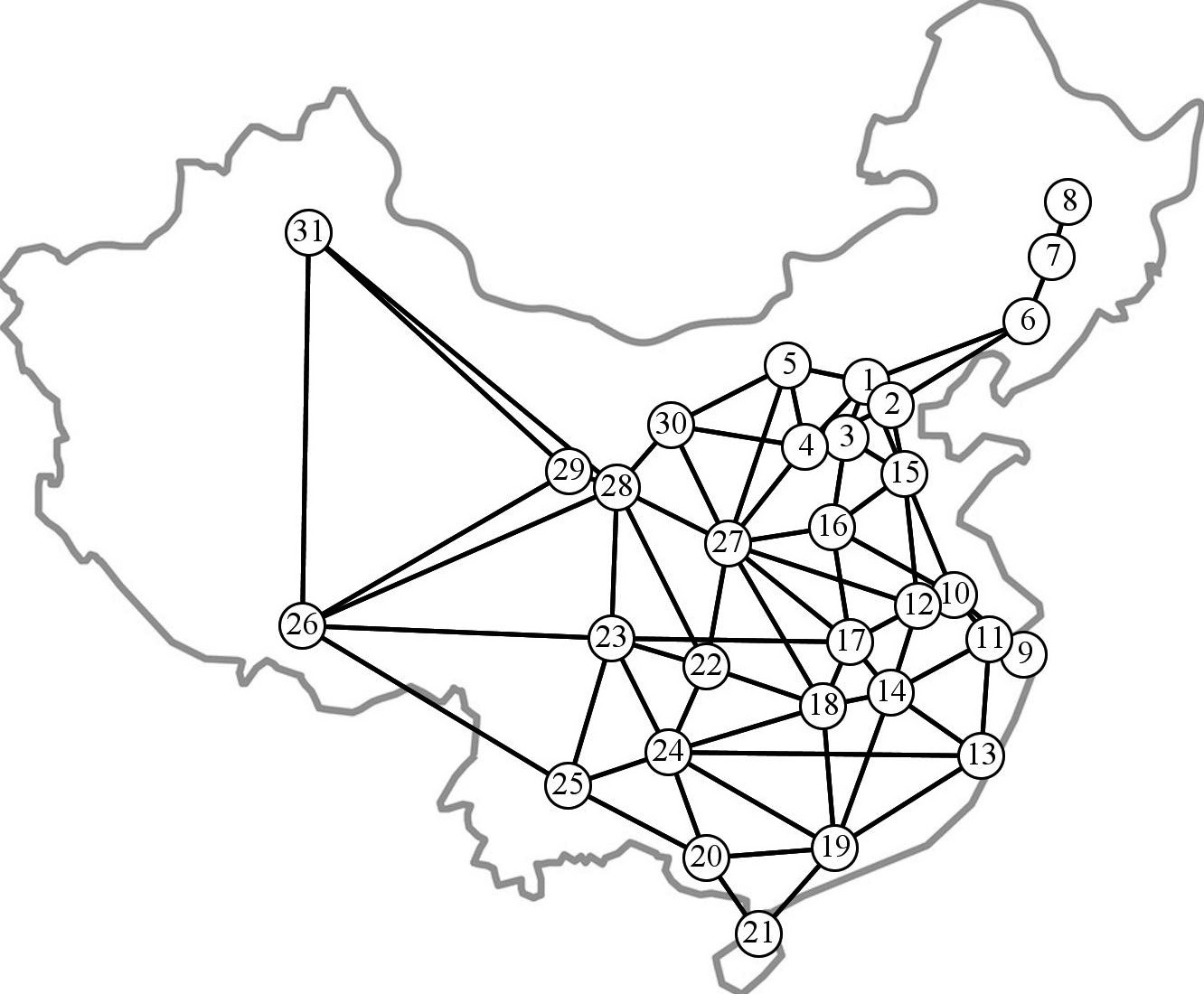}\label{ChinaH}}
\subfigure[Germany]{\includegraphics[scale=0.1]{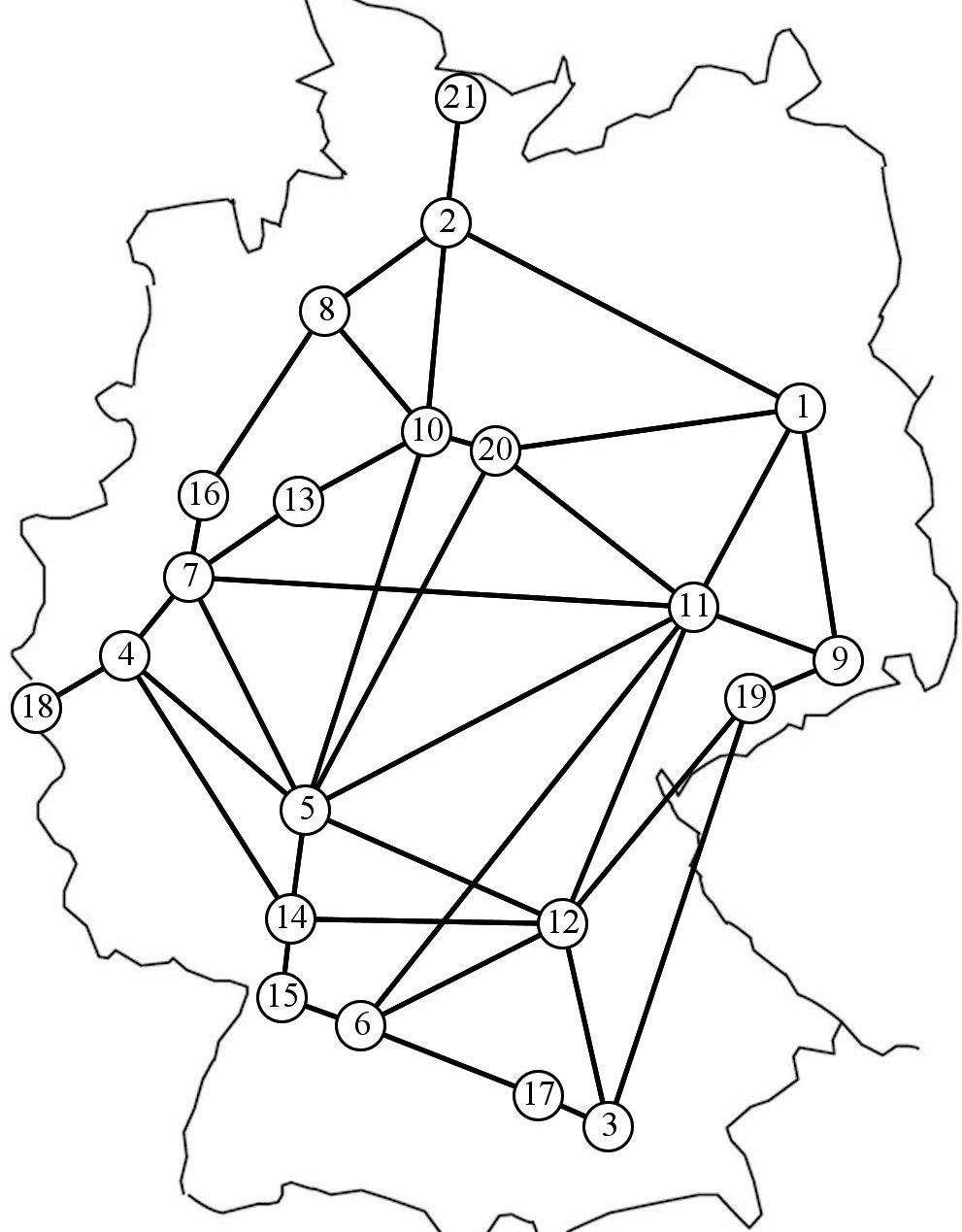}\label{GermanyH}}
\subfigure[Iberia]{\includegraphics[scale=0.1]{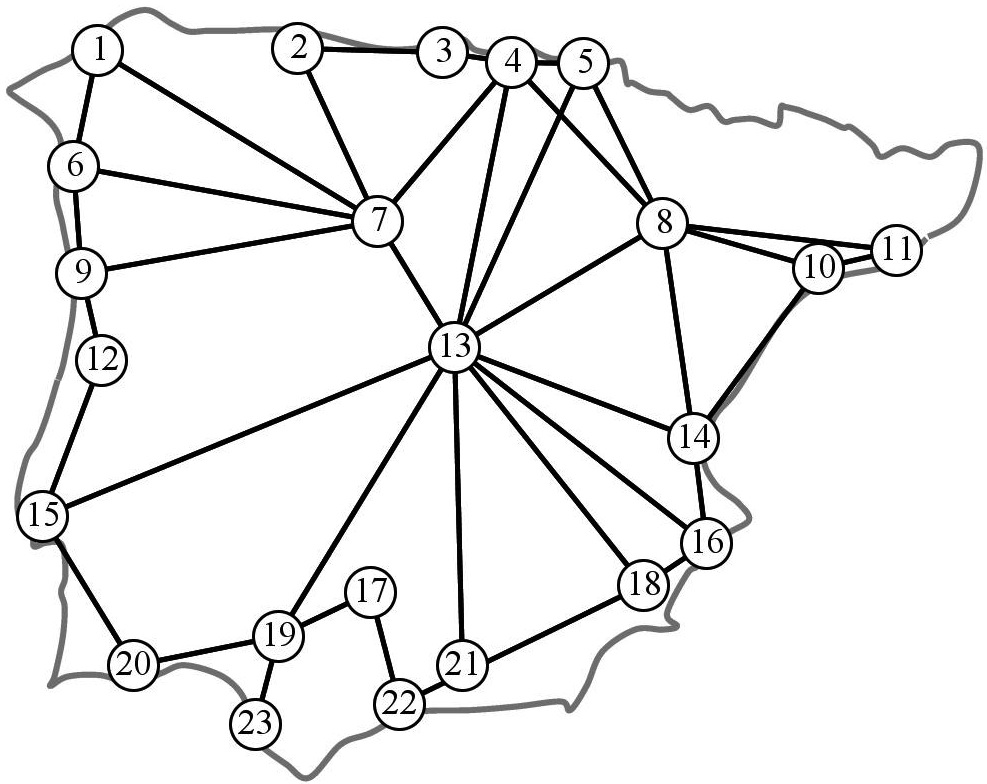}\label{IberiaH}}
\subfigure[Italy]{\includegraphics[scale=0.1]{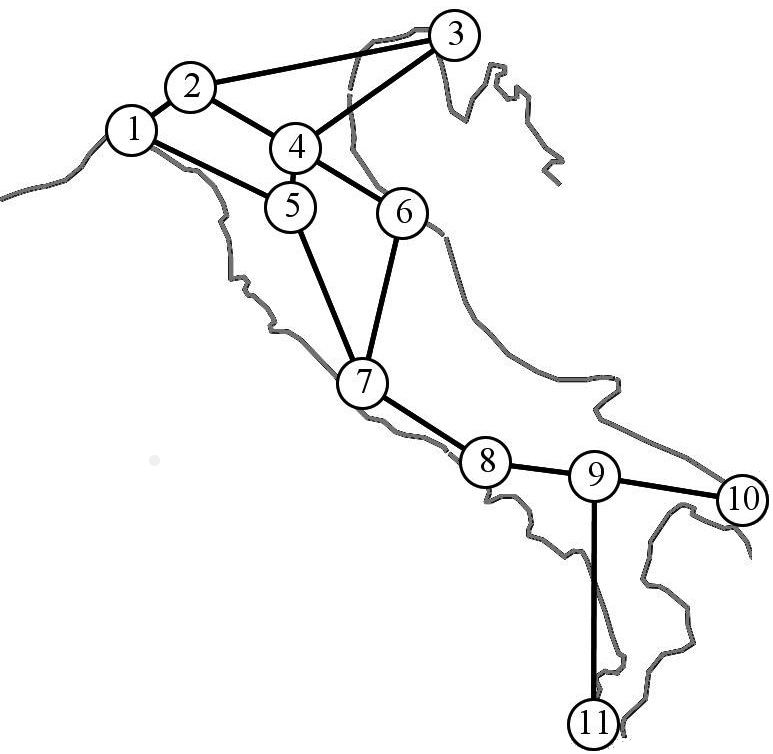}\label{ItalyH}}
\subfigure[Malaysia]{\includegraphics[scale=0.07]{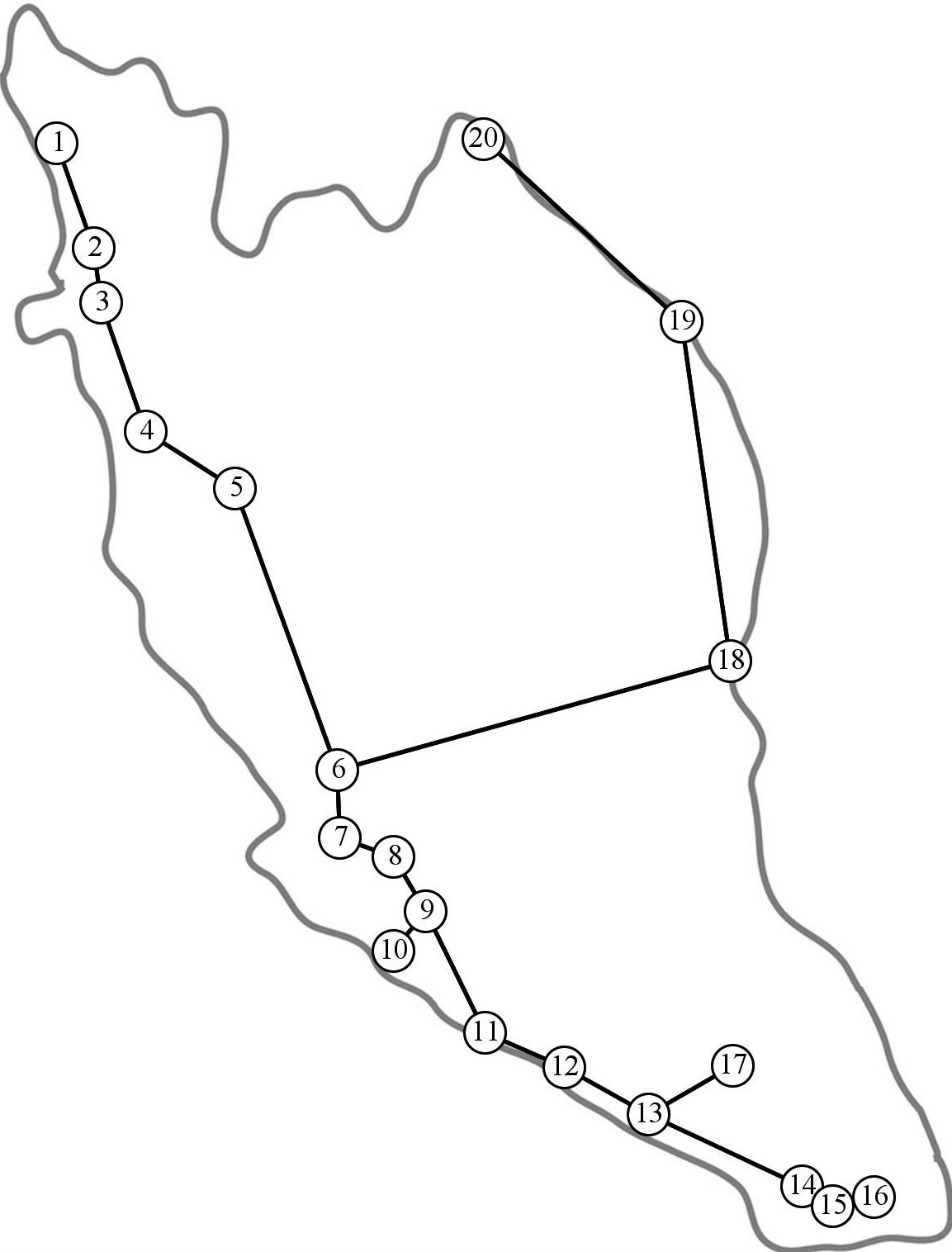}\label{MalaysiaH}}
\subfigure[Mexico]{\includegraphics[scale=0.1]{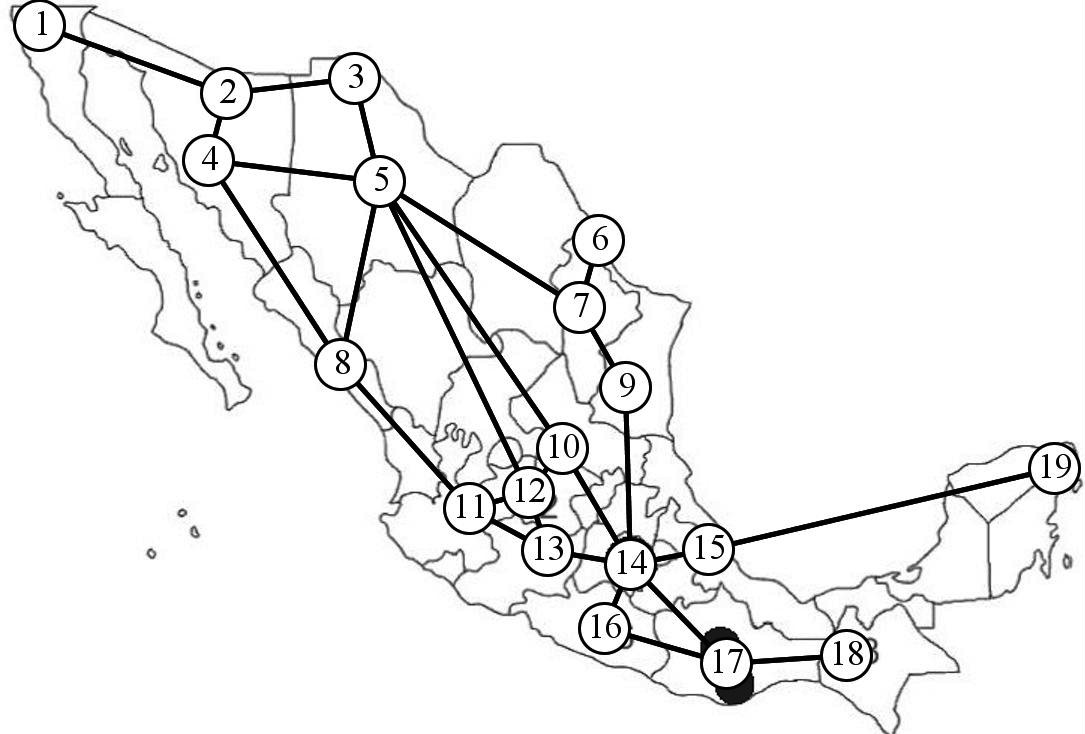}\label{MexicoH}}
\subfigure[the Netherlands]{\includegraphics[scale=0.1]{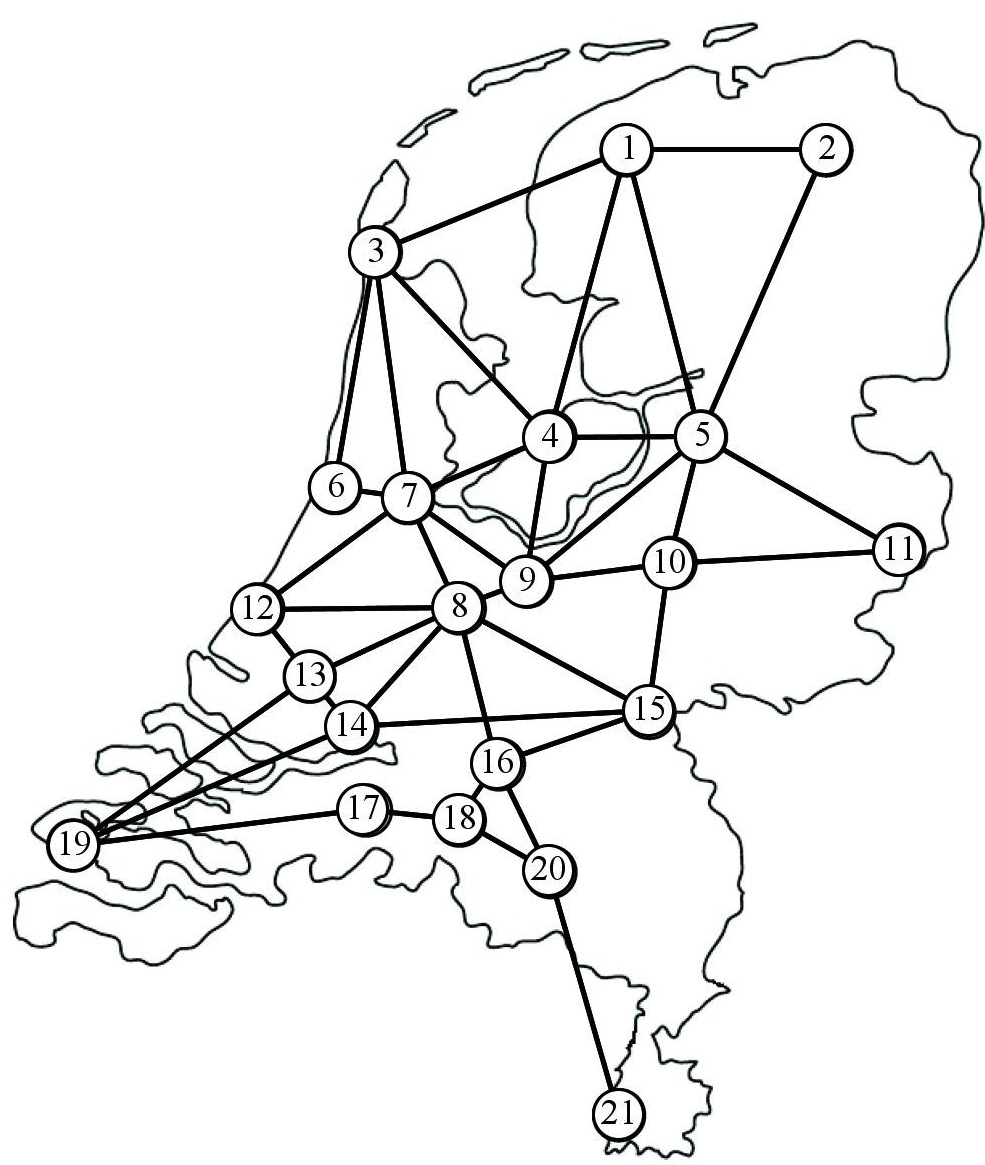}\label{NetherlandsH}} \hspace{0.5cm}
\subfigure[UK]{\includegraphics[scale=0.1]{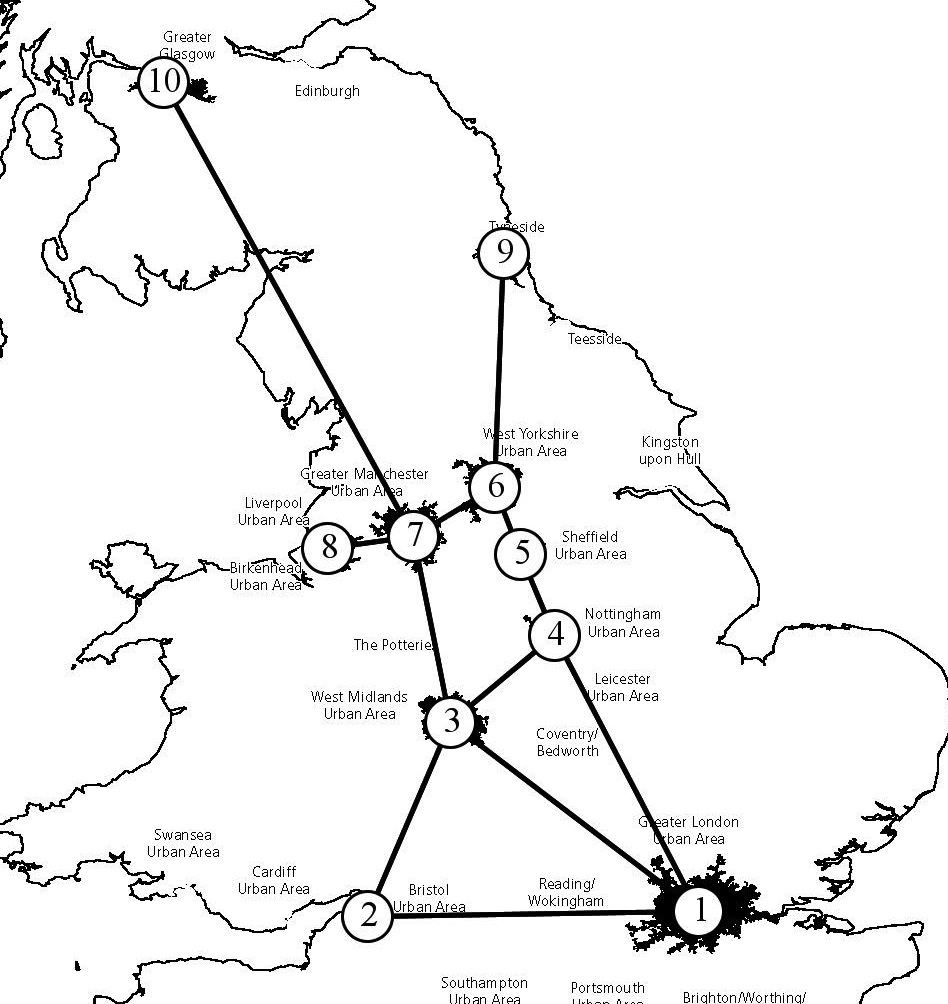}\label{UKH}}
\subfigure[USA]{\includegraphics[scale=0.1]{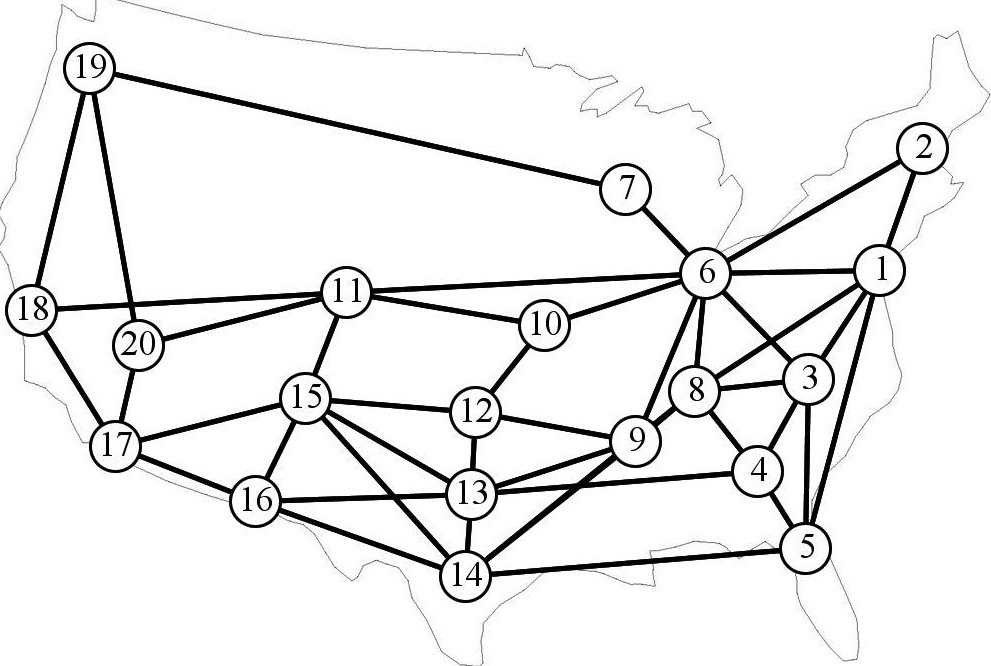}\label{USAsH}}
\caption{Motorway graphs $\mathbf{H}$.}
\label{motorways}
\end{figure}

To compare the Physarum graphs with motorway graphs we construct a motorway graph $\mathbf H$ as follows. Let $\mathbf U$ be a set of urban regions, for any two regions $a$ and $b$ from $\mathbf U$, the nodes $a$ and $b$ are connected by an edge ($a$, $b$) if there is a motorway starting in  $a$ and passing in the vicinity of $b$ and not passing in the vicinity of any other urban area $c \in \mathbf U$. Motorway graphs extracted from maps of motorway/highway/expressway/autobahn networks are shown in Fig.~\ref{motorways}.

Motorway and Physarum graphs were compared directly and using integral measures and indices.  
Let $m$ be a number of edges in motorway graph $\mathbf H$ and $f$ be number of edges in Physarum 
graph $\mathbf P$, and $i$ and $j$ be nodes,  and $M$ and $F$ be adjacency matrices. Direct matching 
between motorway and Physarum graphs is calculated as 
 $\mu = \frac{1}{m} \sum_{ij} \xi (M_{ij}, F_{ij})$, where $\xi(M_{ij}, F_{ij})=1$ if $M_{ij}=F_{ij}$, and 0, otherwise. 
 An economy of matching is calculated as $\epsilon = \frac{\mu}{f}$. Also, we compared 
 the graphs by their average shortest path measured in nodes, average shortest path measured in normalised edge lengths 
 (for each edge $e \in \mathbf{E}$ we normalised its Euclidean length $l(e)$ as  
 $l(e) \leftarrow \frac{l(e)}{max\{l(e'): e' \in \mathbf{E}\}}$, average degrees (sum of degrees of nodes divided by a number of nodes), 
 average edge length (of normalised edges),  diameters (longest shortest path) in nodes and normalised edge lengths, and maximum number of vertex independent cycles (two cycles are independent of each other if they  do not share nodes or edges).
 
 To measure 'compactness' of graphs we calculated average cohesion: let $\overline{d}$ be an average degree of a graph $\mathbf G$ 
 and $\nu_{ij}$ be a number of common neighbours of nodes $i$ and $j$, and $d_i$ is a degree of node $i$, then cohesion  
 $\kappa_{ij}$ between nodes $i$ and $j$ is calculated as $\kappa_{ij}=\frac{\nu_{ij}}{d_i + d_j}$. Three topological indices were 
 calculated: Harary index~\cite{plavsic_1993}, $\Pi$-index~\cite{Ducruet_2012}, and  Randi\'{c} index~\cite{Randic_1975}.

The Harary index is calculated as follows: $H = \frac{1}{2} \sum_{ij} \chi(D_{ij})$, where $D$ is a graph distance matrix, where $D_{ij}$ is 
a length of a shortest path (in normalised edge lengths) between $i$ and $j$, and $\chi(D_{ij})= D^{-1}_{ij}$ if $i \neq j$ and 0, otherwise.
The $\Pi$-index shows a relationship between the total length of the graph $L(\mathbf{G})$ and the distance along its diameter 
$D(d)$~\cite{Ducruet_2012}: $\Pi=\frac{L(G)}{D(d)}$. The Randi\'{c} index~\cite{Randic_1975} is calculated as 
 $R=\sum_{ij} C_{ij}*(\frac{1}{\sqrt{(d_i*d_j)}})$,  where $C_{ij}$ is an adjacency matrix, $C=M$ or $C=F$.

\section{Results}

\subsubsection{Matching and Economy}

Top three regions in the best matches $\mu$ between motorway and Physarum graphs are Malaysia, Italy and Canada and top three most 
economically $\epsilon$ matched are Italy, Brazil and UK (Fig.~\ref{MatchingEconomy}). 

\begin{finding}
Let $C_1 \lhd C_2$ if $\mu(C_1) < \mu(C_2)$, then regions can be arranged in the following hierarchy of absolute Physarum matching:
 USA $\lhd$ Brazil $\lhd$ \{Mexico, Iberia\} $\lhd$ Australia $\lhd$ \{Germany, UK \} $\lhd$ \{Africa, the Netherlands \} $\lhd$ \{China, Belgium \} 
 $\lhd$ Canada $\lhd$ Italy $\lhd$ Malaysia.
 \label{findingAbsoluteMatching}
 \end{finding} 
 
 We can consider a product of matching to economy $\omega= \mu \cdot \epsilon$  as a rough parameter for estimating 'slime-optimality' of motorways approximation. By values of $\omega$  regions can be arranged in the descending slime-optimality as follow
 (exact values of $\omega$ are in brackets): 
\begin{enumerate}
\item Italy (0.85)
\item Malaysia (0.76)
\item Canada (0.67), China (0.65), Belgium (0.6), Africa (0.6), UK (0.59)
\item Netherlands (0.49), Germany (0.48), Australia (0.47), Brazil (0.44), Mexico (0.42), USA (0.4)
\item Iberia (0.34)
\end{enumerate}

\begin{figure}[!tbp]
\centering
\subfigure[Average degrees]{\includegraphics[width=0.49\textwidth]{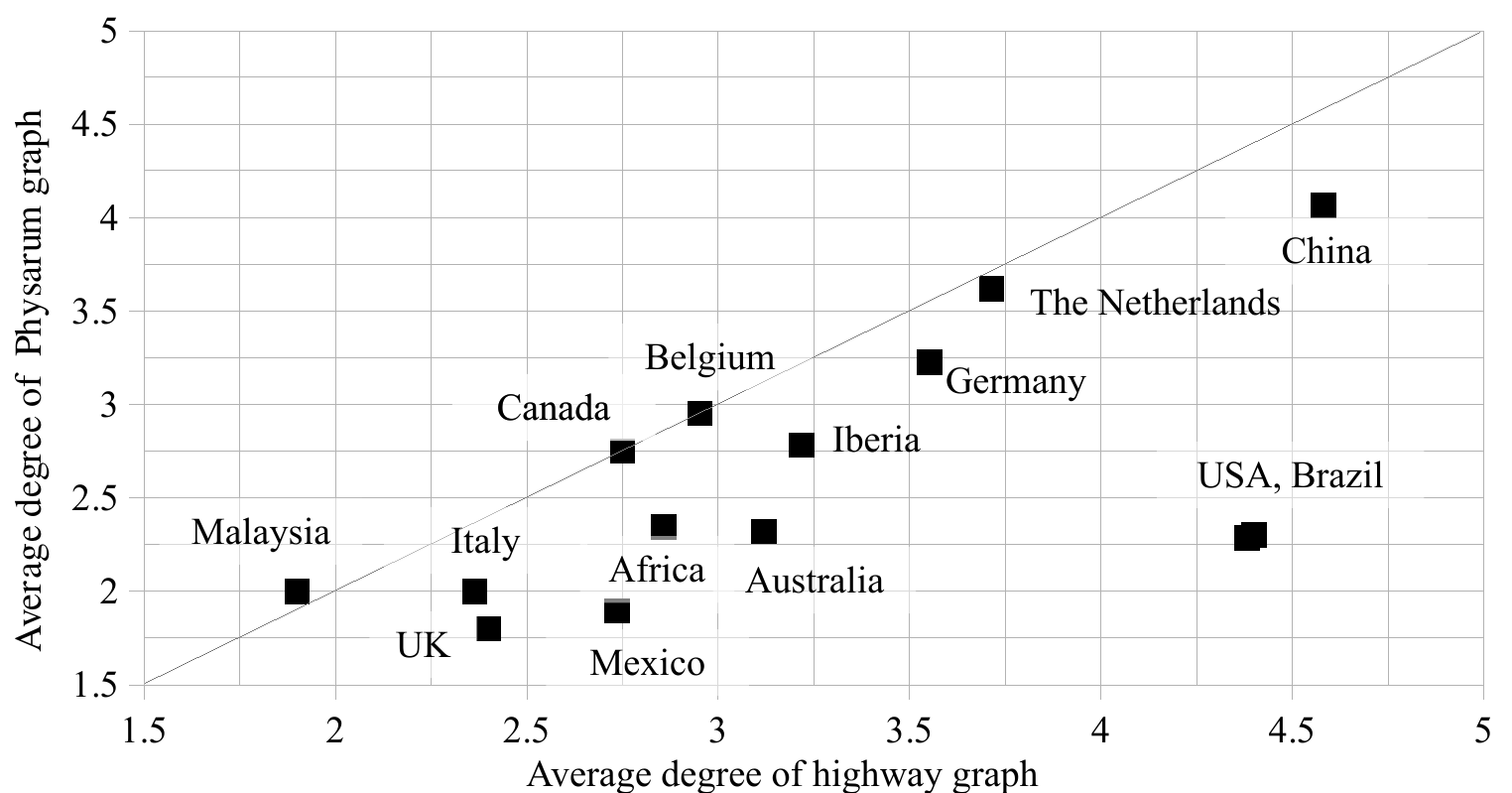}\label{avdegrees}}
\subfigure[Maximum number of independent cycles]{\includegraphics[width=0.49\textwidth]{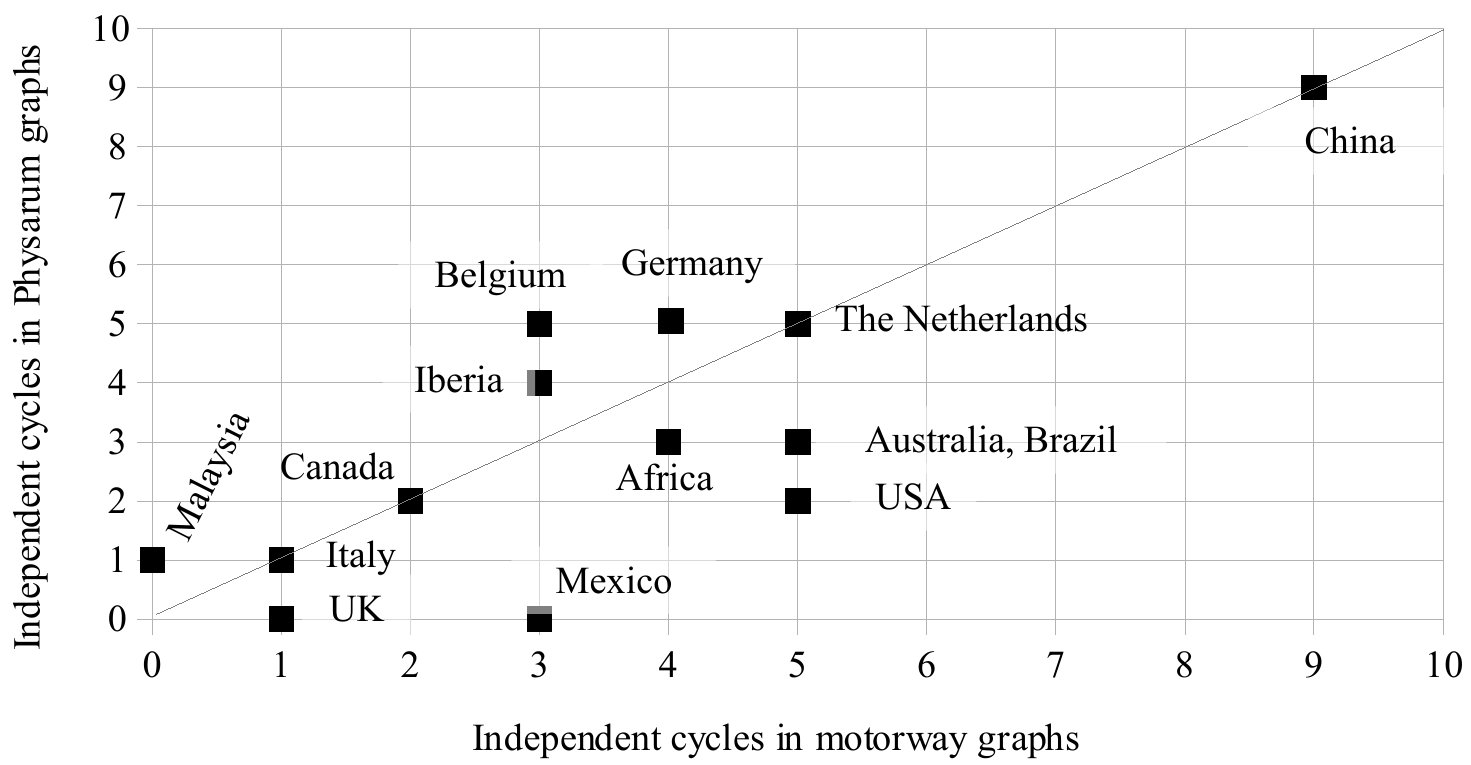}\label{cycles}}
\subfigure[Average edge length]{\includegraphics[width=0.49\textwidth]{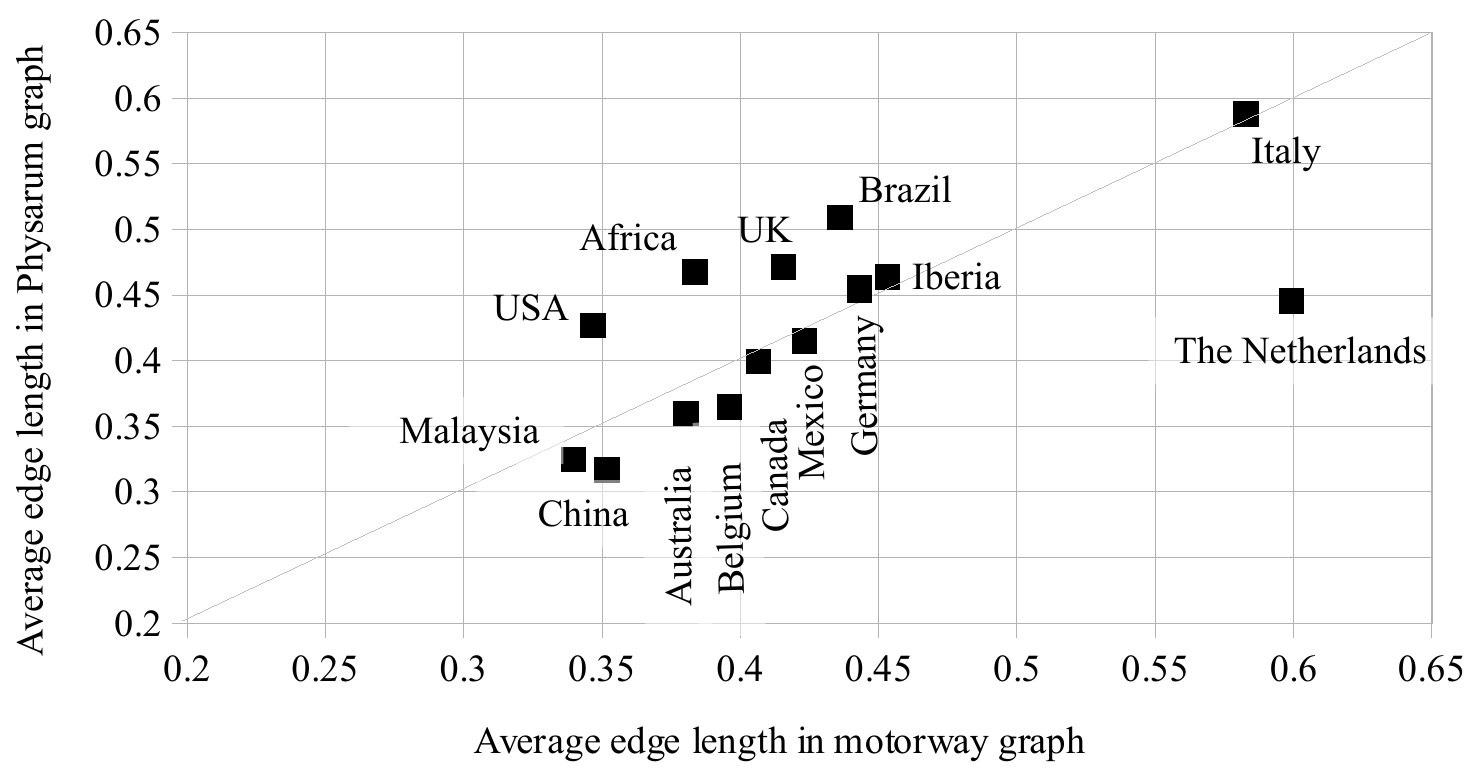}\label{avlink}}
\subfigure[Average shortest path in nodes]{\includegraphics[width=0.49\textwidth]{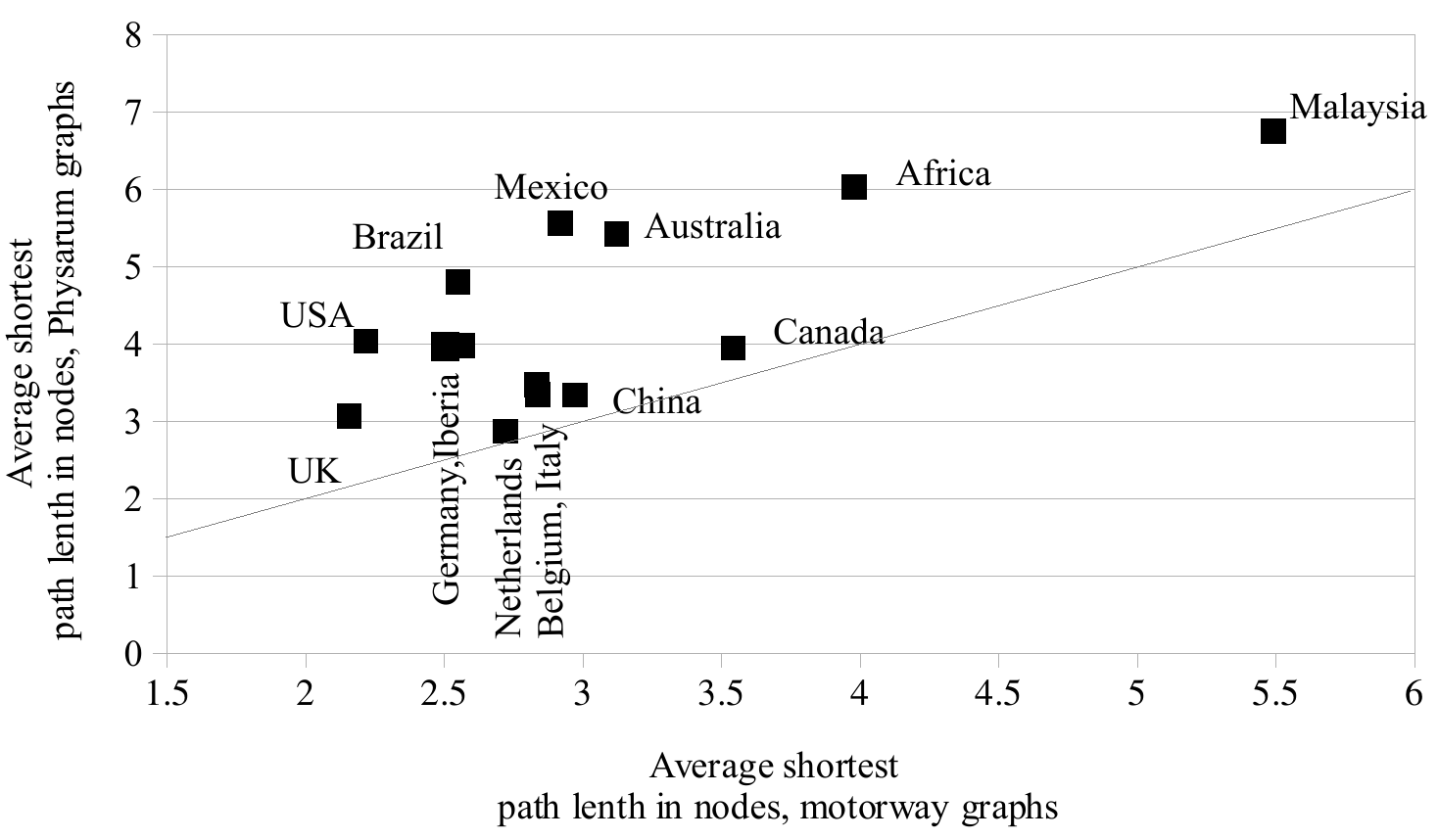}\label{averageSPnodes}}
\subfigure[Average shortest path]{\includegraphics[width=0.49\textwidth]{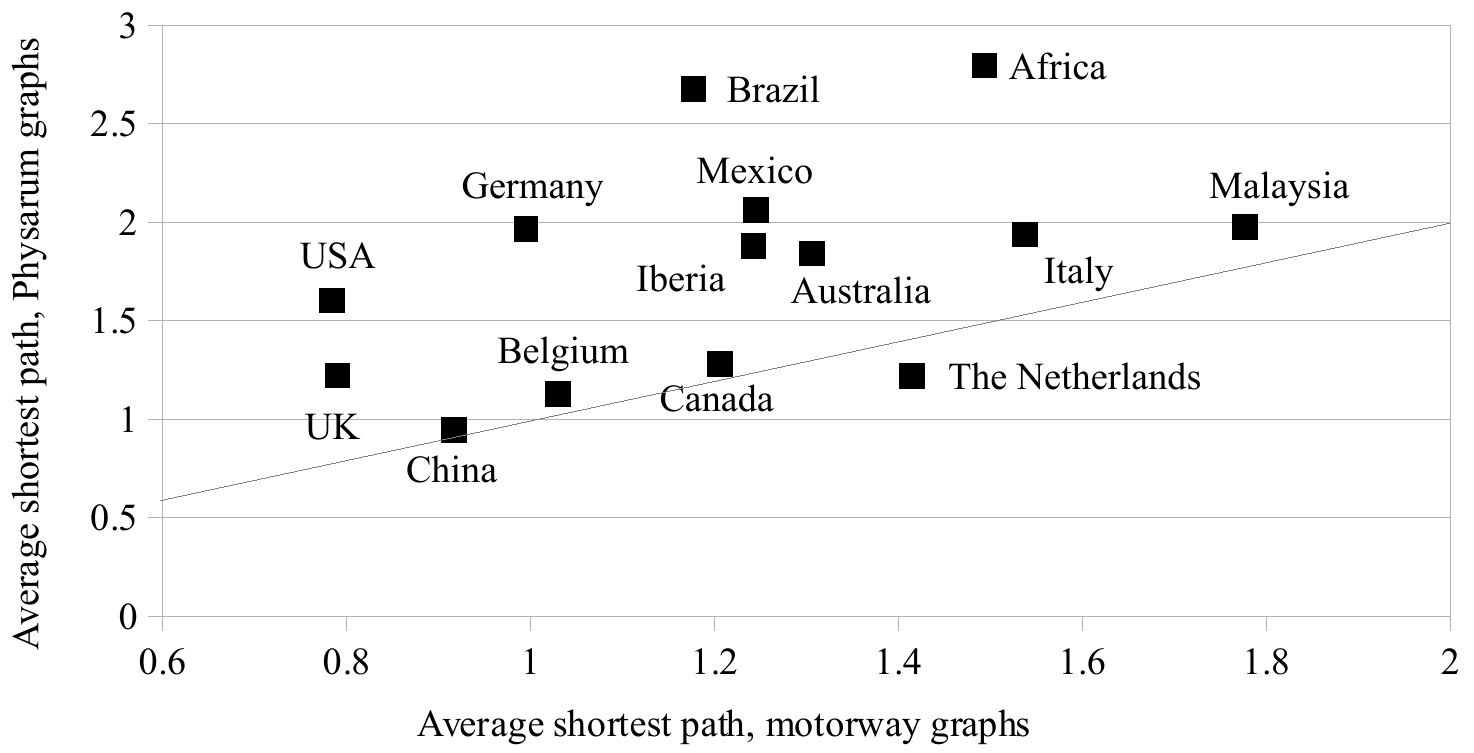}\label{averageSPreal}}
\subfigure[Diameter in nodes]{\includegraphics[width=0.49\textwidth]{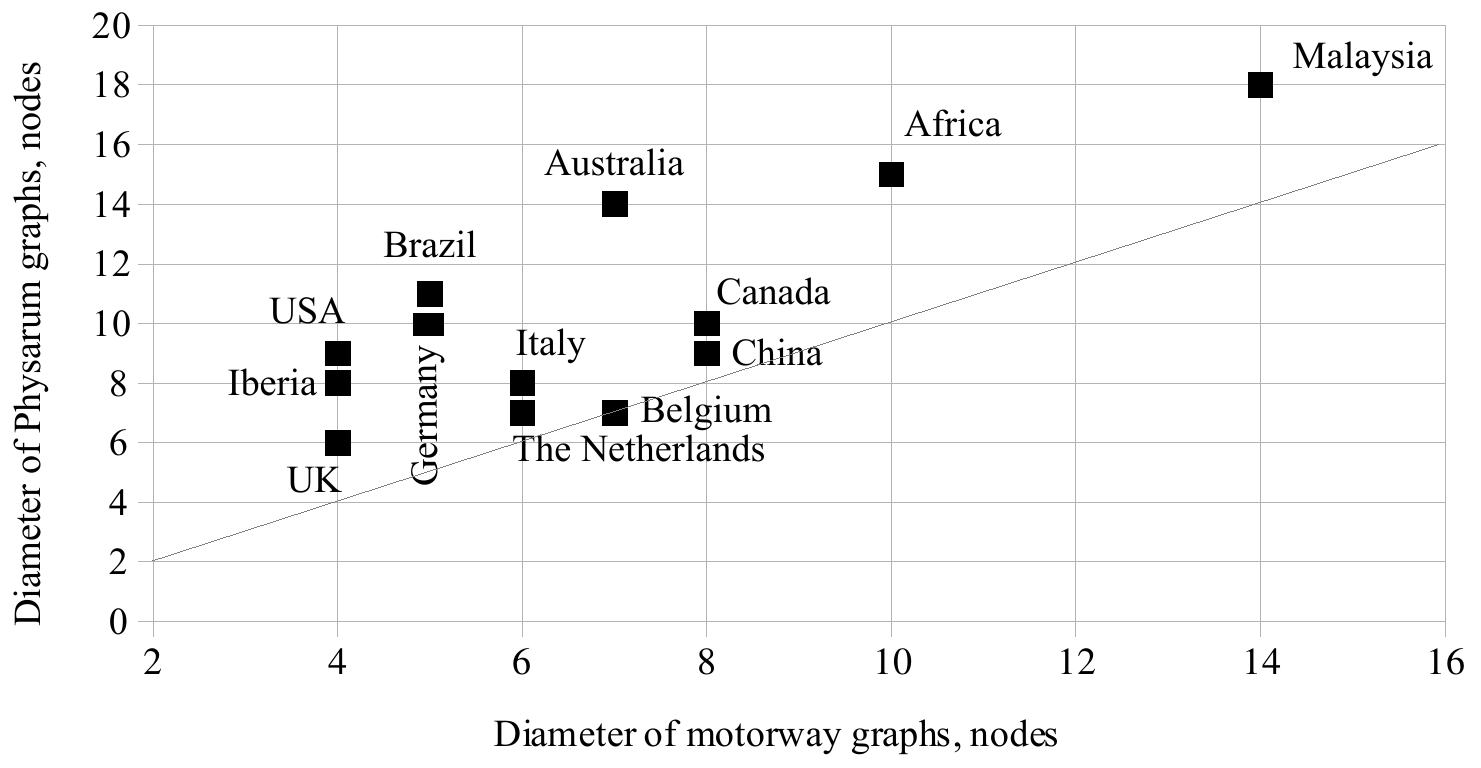}\label{diameterInNodes}}
\subfigure[Diameter]{\includegraphics[width=0.49\textwidth]{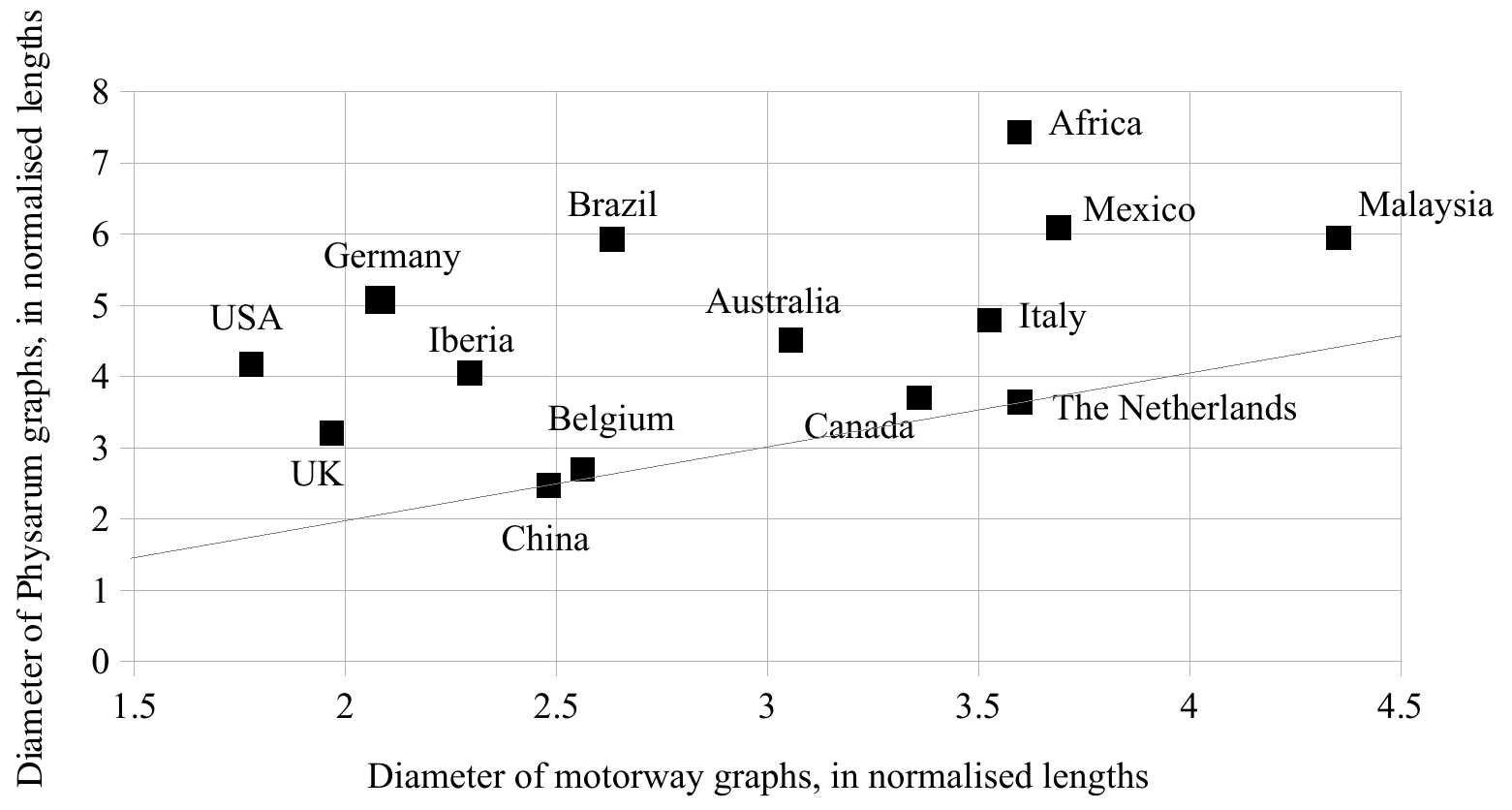}\label{diameterInLengths}}
\caption{
(a)~Average degrees.
(b)~Maximum number of vertex independent cycles.
(c)~Average edge link.
(d)~Average length of a shortest path in nodes.
(e)~Average length of a shortest path in real values.
(f)~Diameter in nodes.
(g)~Diameter in real values.
}
\label{parameters1}
\end{figure}

\subsubsection{Average degrees}

Averages degrees of motorway graphs are usually lower than degrees of the corresponding 
Physarum graphs (Fig.~\ref{avdegrees}). This is particularly visible with USA and Brazil motorway graphs, which 
average degrees are nearly 4.5 while Physarum graphs approximating the motorways have twice less average degree.
Belgium, Canada and Malaysia show almost perfect match between Physarum and motorway graphs in average degrees 
(Fig.~\ref{avdegrees}).

\subsubsection{Maximum number of independent cycles}

Motorway and Physarum graph of China, the Netherlands, Canada  and Italy have the maximum number of independent cycles (Fig.~\ref{cycles}) 
and show a good match. Other regions can be subdivided on two groups: 
\begin{itemize}
\item number of independent cycles is higher in motorway graphs: Africa, Australia, Brazil, Mexico, UK, USA
\item number of independent cycles is higher in Physarum graphs: Belgium, Berlin, Canada, Germany, Iberia and Malaysia.
\end{itemize}
The maximum number of independent cycles may characterise two properties of transport networks: fault-tolerance (more cycles indicate 
more chances for a transported objects to avoid faulty impassable links, sites of accidents, and jams) and locality (distant links 
increase chances of two cycles of sharing a node).
Thus, we can propose that China motorway network is most fault-tolerant and locally connected, while transport networks in Canada, Italy, Malaysia 
and UK could be sensitive to disasters and overloads.

\subsubsection{Average edge length}

The closest match between Physarum and motorway graphs in average edge length is observed for Italy, Iberia, Germany, Mexico, Canada and Malaysia
(Fig.~\ref{avlink}). Highest mismatch is shown by the Netherlands (edges of motorway graph are longer than edges of Physarum graph), 
and Brazil, Africa, and USA (edges of Physarum are longer, in average, than of motorway graphs). 

\subsubsection{Average shortest paths}

The average length of a shortest path between nodes shows little match between Physarum and 
motorway graphs (Figs.~\ref{averageSPnodes} and \ref{averageSPreal}). Usually Physarum graphs 
exhibit 1.5-2 times longer average shortest paths, this varies however between regions. Malaysia and Africa 
are the regions with longest average shortest paths,  measured in nodes, in motorway and Physarum 
graphs (Figs.~\ref{averageSPnodes}). When shortest paths are measured in normalised edge length, 
Africa and Brazil have longest average shortest paths in Physarum graphs, and Malaysia and Italy in motorways 
graphs. Countries which show closest match between Physarum and motorway graphs 
--- in average shortest path, measured in nodes (Figs.~\ref{averageSPnodes}) --- are Canada, China, the Netherlands; and --- in average
shortest path measured in normalised lengths (Fig.~\ref{averageSPreal}) --- are Belgium, Canada, China.

\subsubsection{Diameters}

Being the longest shortest path, the diameter shows even less matching 
between Physarum  and motorway graphs (Figs.~\ref{diameterInNodes} 
and \ref{diameterInLengths}) than average shortest path matching.
Physarum graphs match motorways in diameter measured in nodes  for 
Belgium, and  in diameter measured in normalised lengths  for Belgium, China, and the Netherlands.

\begin{figure}[!tbp]
\centering
\subfigure[Average cohesion]{\includegraphics[width=0.49\textwidth]{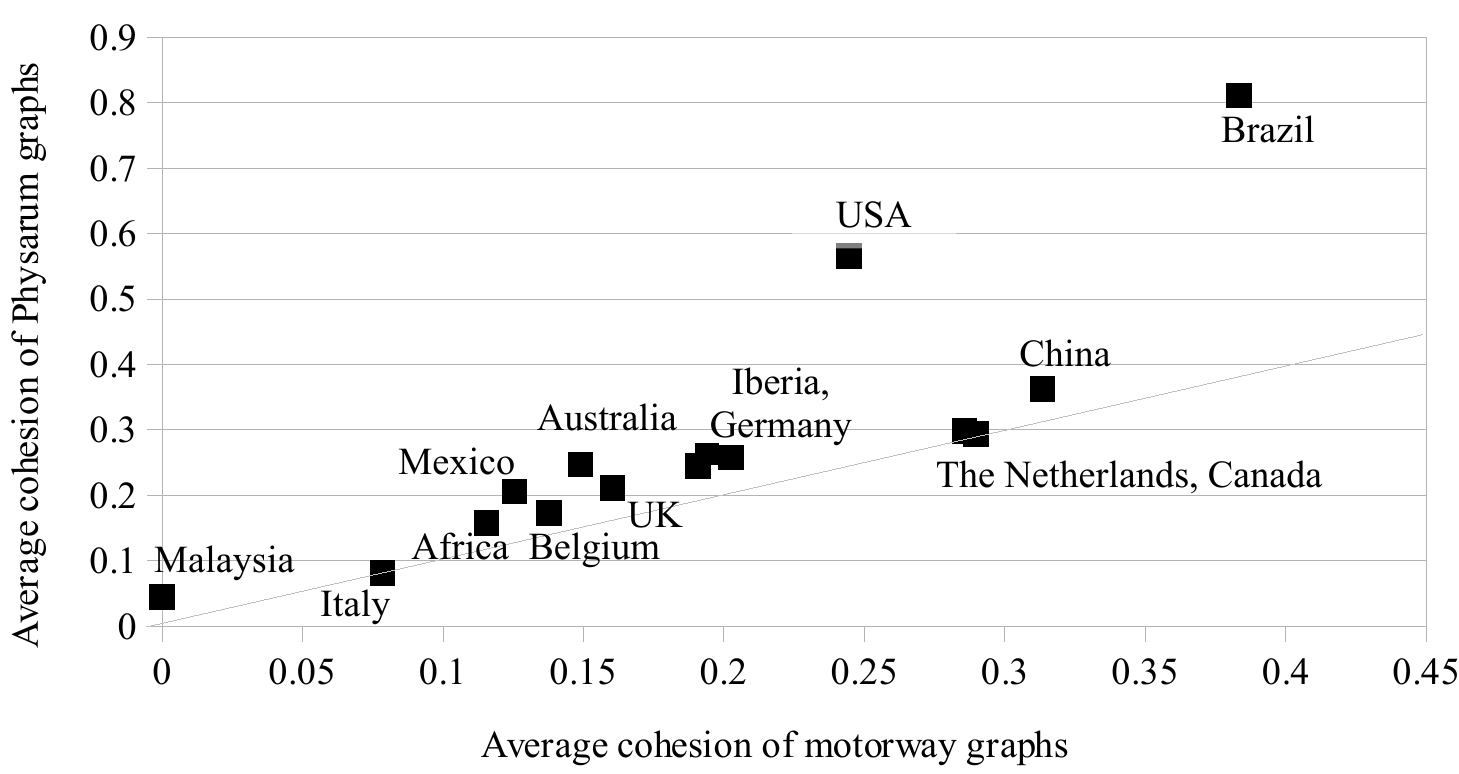}\label{cohesion}}
\subfigure[Harary index]{\includegraphics[width=0.49\textwidth]{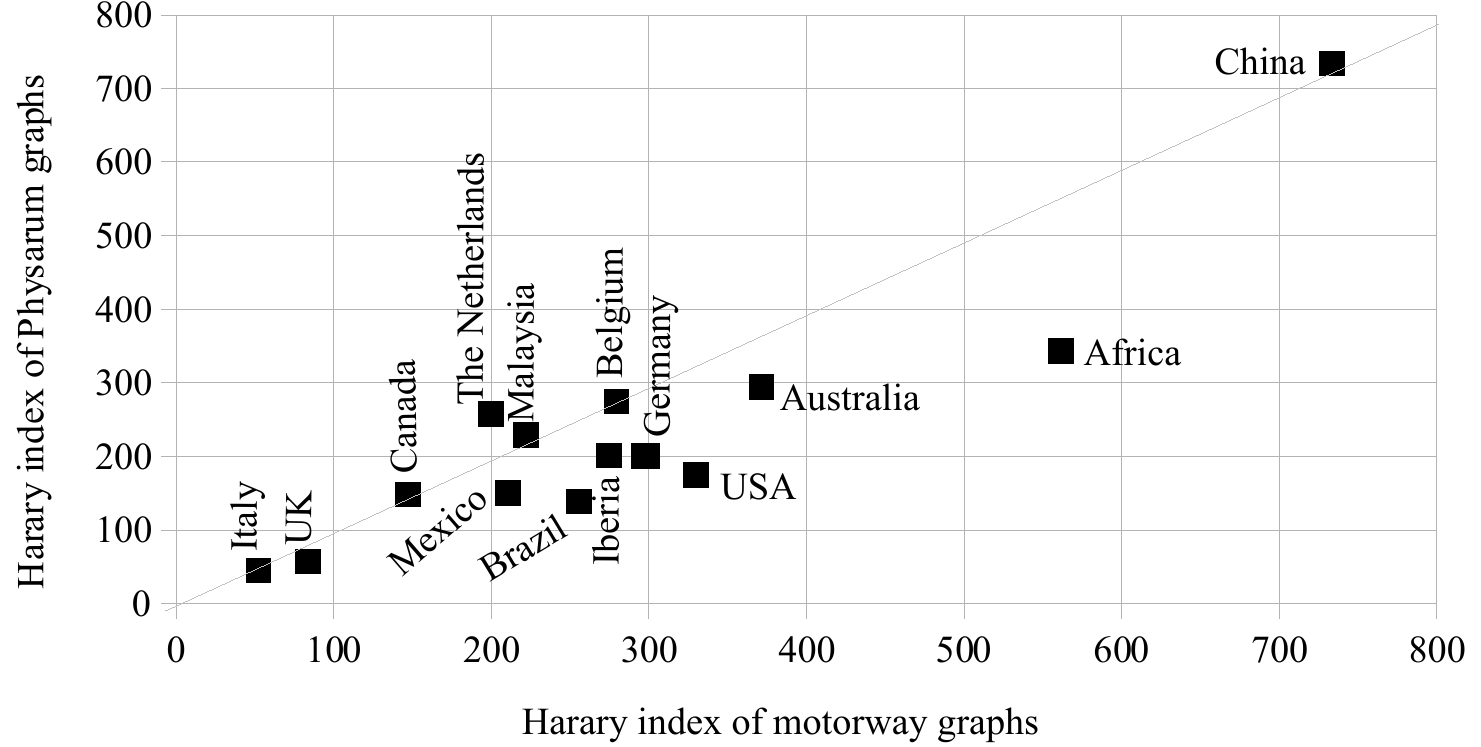}\label{harary}}
\subfigure[$\Pi$-index]{\includegraphics[width=0.49\textwidth]{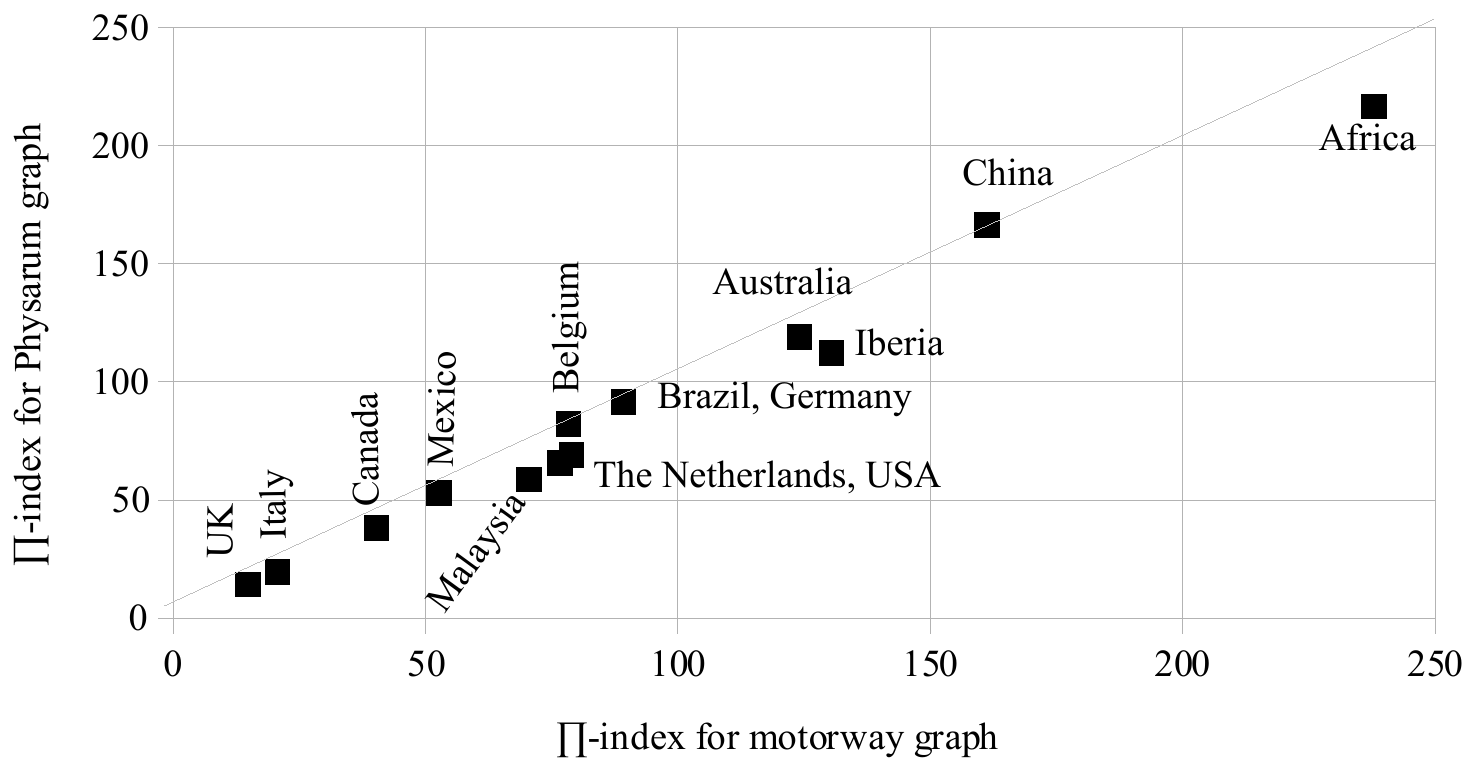}\label{PI_index}}
\subfigure[Randi\'{c} index]{\includegraphics[width=0.49\textwidth]{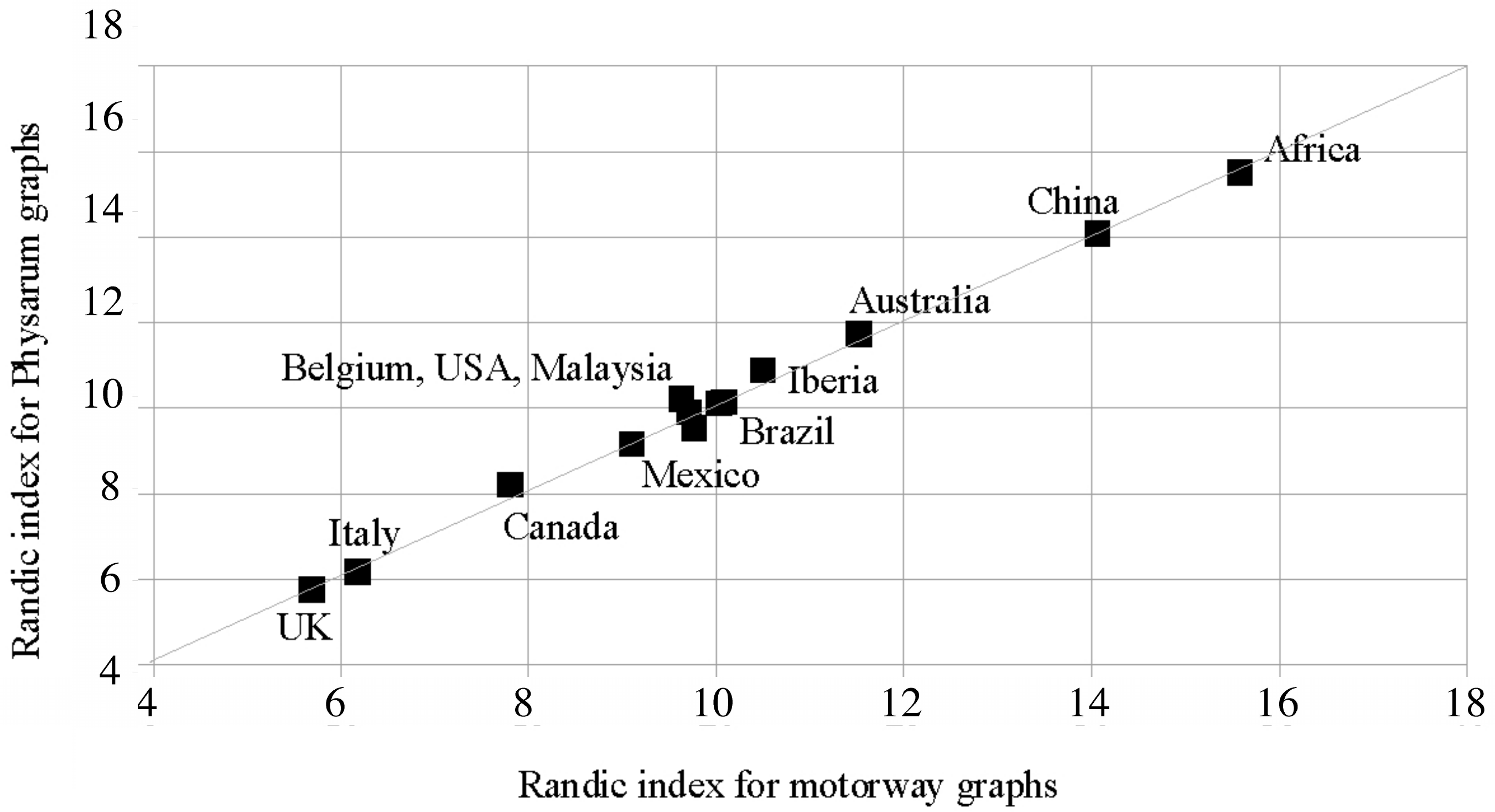}\label{RadicIndex}}
\caption{
(c)~Average cohesion.
(c)~the Harary index.
(d)~the $\Pi$ index.
(e)~the Randi\'{c} index.
}
\label{parameters2}
\end{figure}

\subsubsection{Cohesion}

For most regions considered, average cohesion of Physarum graphs is typically higher
than of motorway graphs (Fig.~\ref{cohesion}). The difference is particularly strong for
Brazil and USA, e.g. average cohesion of Brazilian motorway network is 0.8 while of the corresponding 
Physarum graphs it is 0.4. There is a match between Physarum and motorway graphs of Canada, Italy, and the Netherlands. 
The top three entries with highest cohesion of motorway graph are Brazil, Canada, China, and the Netherlands, and the top three 
regions with highest cohesion of Physarum graphs are  Brazil, China and USA. Cohesion of each edge in the complete graph 
is $\frac{n-2}{2(n-1)} \sim 0.44$, this limit is nearly approached by $\mathbf{H}$(Brazil) (Fig.~\ref{cohesion}). Motorway and, 
especially, Physarum,  graphs of Malaysia and Italy show minimal cohesion, because cohesion is zero in chains and graphs with 
cycles over three nodes.

\begin{finding}
Physarum matches motorway network of Canada, Italy and the Netherlands in terms of compactness, local densities and fault-tolerance of transport
networks.
\end{finding}

Average cohesion is an indicator of compactness~\cite{egghe_2003}. A sub-set of the graph with high cohesion remains connected 
even when some edges are removed, thus cohesion may characterises stability~\cite{seidman_1983} or even fault-tolerance of graphs.
The cohesion of a node  is the minimum number of edges whose deletion makes the node a cut node of the resulting graph~\cite{Ringelsen_1983}, thus the cohesion is used to characterise a local density of sub-graphs~\cite{tibely_2012}, and it is related to centrality~\cite{Borgatti_2006}, and 
statistical properties of connectivity of graphs~\cite{Tainiter_1975}.

\subsubsection{The Harary index}

The Harary index~\cite{plavsic_1993} is well know for its predictive properties in chemistry~\cite{estrada_1997a,estrada_1997b}, and the index 
is based on  `the chemists' intuitive expectation that distant sites in a structure should influence 
each other less than the near site'~\cite{lucic_2002}. This is probably not the case with slime mould and man-made motorway networks. 
Only four out fourteen regions satisfy the relation $|1-\frac{\eta(P)}{\eta(H)}| \leq 0.1$: China, Canada, Belgium and Malaysia, 
where $\eta$ is the Harary index. Physarum poorly approximates motorways, in terms of the Harary index, in Brazil and USA (Fig.~\ref{harary}).

\subsubsection{The $\Pi$-index}

Five regions show over 0.1 mismatch, $|1-\frac{\Pi(P)}{\Pi(H)}| > 0.1$, between the $\Pi$-indices of Physarum and 
motorway graphs: Germany, Iberia, the Netherlands, Malaysia, and USA (Fig.~\ref{PI_index}). This result is quite interesting. 
Recall, that the $\Pi$-index is  a ratio of a total length of all normalised edges of a graph to a distance along the graph's diameter. 
Physarum graphs neither match motorways in diameters (Figs.~\ref{diameterInNodes} and \ref{diameterInLengths}),
nor do we witness a good match in average edge length, or shortest paths (Figs.~\ref{avlink},~\ref{averageSPnodes}, 
and~\ref{averageSPreal}). However, when these factors are considered in proportion the match between the graphs occurs.

\subsubsection{The Randi\'{c} index}

\begin{finding}
\label{randic}
Physarum perfectly approximates motorway networks in terms of the Randi\'{c} index.
\end{finding}

\begin{table}
\caption{Values of the Randi\'{c} index.}
\centering
\begin{tabular}{l|ccccc}
Country	&	$R(\mathbf{H})$	&	$R(\mathbf{P})$	&	$1-\frac{R(\mathbf{H})}{R(\mathbf{P})}$	&	$\frac{n}{2}$	&	\\ \hline
Africa 	&	16.98	&	16.88	&	-0.006	&	17.5	&	\\ 
Australia	&	11.9	&	12.17	&	0.022	&	12.5	&	\\ 
Belgium	&	9.53	&	10.27	&	0.071	&	10.5	&	\\ 
Brazil	&	10.04	&	10.14	&	0.01	&	10.5	&	\\ 
Canada	&	7.25	&	7.76	&	0.066	&	8	&	\\ 
China	&	15.08	&	15.1	&	0.002	&	15.5	&	\\ 
Germany	&	9.94	&	10.15	&	0.02	&	10.5	&	\\ 
Iberia	&	10.62	&	11.11	&	0.045	&	11.5	&	\\ 
Italy	&	5.22	&	5.22	&	0	&	5.5	&	\\ 
Malaysia	&	9.63	&	9.88	&	0.025	&	10	&	\\ 
Mexico	&	8.87	&	8.95	&	0.009	&	9.5	&	\\ 
Netherlands	&	10.09	&	10.17	&	0.008	&	10.5	&	\\ 
UK	&	4.61	&	4.7	&	0.021	&	5	&	\\ 
USA	&	9.71	&	9.41	&	-0.031	&	10	&	\\ 

\end{tabular}
\label{tabRandic}
\end{table}

The Randi\'{c} index $R$ shows impeccable match between Physarum and motorway graphs 
(Fig.~\ref{RadicIndex} and Tab.~\ref{tabRandic}). The largest value $1-\frac{R(\mathbf{H})}{R(\mathbf{P})}=0.07$ 
is for Belgium motorway graph, and corresponding Physarum graph.  
Star-graph has a minimum the Randi\'{c} index $\sqrt{n-1}$~\cite{bolobas_1998}. For an arbitrary 
graph $\mathbf{G}$  the boundaries are $\sqrt{n-1} \leq R(\mathbf{G}) \leq \frac{n}{2}$. As we can see in 
Tab.~\ref{tabRandic}, indices for all motorway and Physarum graphs are very close to upper boundary. The highest values
for motorways are in Italy and USA, and for Physarum graphs are in Belgium, Malaysia and Canada.  

 The Randi\'{c} index $R$ (originally called by Milan Randi\'{c} as molecular branching index)~\cite{Randic_1975} 
 characterises relationships between structure, property and activity of molecular components~\cite{estrada_2001}. 
 The index  relates to diameter~\cite{dvorak_2011} and is actually the upper boundary of diameter~\cite{yang_2011}. It also
 relates to chromatic numbers of graphs and eigenvalues of adjacency matrices~\cite{li_2008}.
 There are proven linear relations between the  Randi\'{c} index and molecular polarisability, cavity surface areas calculated for water solubility of alcohols and hydrocarbons,  biological potencies of  anaesthetics~\cite{kier_1975}, water solubility and boiling
 point~\cite{hall_1975} and even bio-concentration factor of hazardous chemicals~\cite{sablijc}. Estrada~\cite{estrada_2002}
 suggested the following structural interpretation: the Randi\'{c} index is proportional to an area of molecular accessibility, i.e. area 
 'exposed' to outside environment. Or we can say that the index is inversely proportional to areas of overlapping between spheres of specified radius enclosing the nodes. The more overlapping the less is Randi\'{c} index. In terms of transport networks, we can interpret external accessibility as transport inaccessibility, proportional to areas of country not served by existing motorway links. 
 
 \begin{finding}
 Physarum well approximates motorway graphs in terms of transport accessibility.
 \end{finding} 
   
 Along the above discourse we can speculate that UK, Italy and Canada (first three regions with the smallest Randi\'{c} indices)  have 
 better transport coverage of their territories than Africa, China and Australia (top three regions with the highest Randi\'{c} indices).

 \subsubsection{Extremal regions}

\begin{table}
\caption{Geographical regions with extremal values of measures over motorway graphs $\mathbf{H}$ and Physarum graphs $\mathbf{P}$.}
\begin{tabular}{l|llll}
Measure $\mu$ 		& $\max \mu(\mathbf{H})$ &  $\min \mu(\mathbf{H})$ 	&  $\max \mu(\mathbf{P})$ &  $\min \mu(\mathbf{P})$ \\ \hline
Average degree 		& China 		&	 Malaysia		 &  China 						& UK   \\
Number of independent cycles & China		&        Malaysia 		 &  China 						& UK, Mexico \\
Average cohesion 		& Brazil		&        Malaysia		 &  Brazil						&  Malaysia \\
Average edge length 	& the Netherlands	&  Malaysia			&  Italy 						& China   \\
Shortest path, nodes 	& Malaysia 		& UK 				&  Malaysia 						& the Netherlands \\
Shortest path 		& Malaysia 		&  UK, USA  			& Africa 						& China \\
Harary index 		& China 		& Italy 			& China 						& Italy  \\
$\Pi$-index 			& Africa  		& UK  				& Africa 						& UK \\
Randi\'{c} index 		& Africa  		& UK  				& Africa 						& UK \\
Diameter, nodes 		& Malaysia 		& Iberia, UK, USA 		& Malaysia 						& UK\\
Diameter 			& Malaysia 		& USA  			& Africa 						& China \\
\label{extremal}
\end{tabular}
\end{table}

We call a region extremal if it displays minimum or maximum values of at least one measure over its motorway or Physarum graphs. 
The extremal regions are listed in Tab.~\ref{extremal}: 
\begin{itemize}
\item Africa shows maximum $\Pi$-index and Randi\'{c} index on both \h an \p, and maximum average 
shortest path and diameter on \p. This might indicate on critical dependencies between geographically close 
urban areas, large territorial spread of transport networks and relatively higher density of urban area along coasts (Fig.~\ref{AfricaH}).
\item Brazil shows maximum average cohesion on \h and \p. This is because Brazil has highest (amongst regions studied)  number of 
locally connected sub-graphs with larges number of dependent cycles (Fig.~\ref{BrazilH}). 
\item China shows maximum average degree, number of independent cycles and Harary index on \h and \p; minimum 
average edge length, average shortest path and diameter on \p. These indicate on high accessibility of major urban area in China, 
and fault-tolerance of Chinese motorways at a large-scale (Fig.~\ref{ChinaH}). The expressway network in China has been developed much recently, known as the national trunk highway system, and it is a high-standard transport system planned by the central government. The system is designed to 
be optimal and many factors were properly taken into account including terrains and landscapes.
\item Iberia has minimum diameter in nodes on \h. The man-made transport network structure resembles a wheel with an 'axle' at Madrid, most major urban areas around coast forming a 'rim'  linked to Madrid by 'spokes' (Fig.~\ref{IberiaH}).  
\item Italy shows minimum Harary index on \h and \p, and maximum average edge length on \p. These are due to tree-like structure of the transport networks and constraining of the urban areas in the prolonged shape of the country (Fig.~\ref{ItalyH}).
\item Malaysia shows maximum shortest path in nodes and diameter in nodes on \h and \p;
minimum average cohesion on \h and \p; 
maximum  average shortest path and diameter on \h;
minimum average degree, number of independent cycles, average edge length on \h. This is because Malaysian transport network does not have
cycles and consists of a chain connecting urban areas along western coast (Fig.~\ref{MalaysiaH}). 
\item Mexico shows  minimum number of independent cycles on \p because the slime mould approximation is a tree, almost a chain with 
few branches (Fig.~\ref{MexicoP}).
\item The Netherlands shows maximum average edge length on \h and minimum average shortest path in nodes on \p, due to relatively compact 
location of urban areas with high density of local transport links (Figs.~\ref{NetherlandsH} and~\ref{NetherlandsP}).
\item UK shows minimum $\Pi$-index and Randi\'{c} index and diameter in nodes on \h and \p; 
minimum average shortest path in nodes, shortest path on \h, and minimum average degree and number 
of independent cycles on \p (Figs.~\ref{UKH} and~\ref{UKP}); 
\item USA shows minimum average shortest path, diameter in nodes and diameter on \h.  These are because the motorway system in USA was built with optimality yet efficiency in mind.
\end{itemize}
 
 \subsubsection{Bio-rationality of measures}

 \begin{table}
 \caption{Matching $M$ between Physarum and motorway graphs for each country and each measure.}
 \centering
 \footnotesize
 \begin{tabular}{p{3cm}||p{1cm}|p{1cm}|p{1cm}|p{1cm}|p{1cm}|p{1cm}|p{1cm}|p{1cm}|p{1cm}|p{1cm}|p{1cm}||p{1cm}|}
 	&	Av \-degree	&	Cycles	&	Co\-hesion	&	Edge length	&	SP in nodes	&	SP	&	Harary-index	&	Pi-index	&	the Randic index	&	Diame\-ter, nodes	&	Diame\-ter	&	Bio-ratio\-nality of motor\-ways	\\ \hline \hline
Africa	&		&		&		&		&		&		&		&		&	1	&		&		&	1	\\
Australia	&		&		&		&		&		&		&		&	1	&	1	&		&		&	2	\\
Belgium	&	1	&		&		&		&		&	1	&	1	&	1	&	1	&	1	&	1	&	7	\\
Brazil	&		&		&	1	&		&		&		&		&	1	&	1	&		&		&	3	\\
Canada	&	1	&	1	&		&	1	&		&	1	&	1	&	1	&	1	&		&		&	7	\\
China	&		&	1	&		&		&		&	1	&	1	&	1	&	1	&		&	1	&	6	\\
Germany	&		&		&		&	1	&		&		&		&	1	&	1	&		&		&	3	\\
Iberia	&		&		&		&	1	&		&		&		&		&	1	&		&		&	2	\\
Italy	&		&	1	&		&	1	&		&		&	1	&	1	&	1	&		&		&	5	\\
Malaysia	&	1	&		&	1	&	1	&		&		&	1	&		&	1	&		&		&	5	\\
Mexico	&		&		&		&	1	&		&		&		&	1	&	1	&		&		&	3	\\
The Netherlands	&	1	&	1	&		&		&		&		&		&		&	1	&		&	1	&	4	\\
UK	&		&		&		&		&		&		&	1	&	1	&	1	&		&		&	3	\\
USA	&		&		&		&		&		&		&		&		&	1	&		&		&	1	\\ \hline
Bio-rationality of measures &	4	&	4	&	2	&	6	&	0	&	3	&	6	&	9	&	14	&	1	&	3	&		\\
\end{tabular}
\label{largetable}
\end{table}
 

In Tab.~\ref{largetable} we provide binary evaluation $M(C,\mu)$ of matching between Physarum \p and motorway \h graphs calculated 
for each country $C$ and measure $\mu$ as follows: $M(C,\mu) = 1$ if $|1 - \frac{\mu(\mathbf{H}(C))}{\mu(\mathbf{P}(C))}| \leq 0.1$ 
and 0, otherwise.  A bio-rationality $\beta$ of a measure $\mu$ is a number of regions $C$ for which $\mu(\mathbf{H}(C))=\mu(\mathbf{P}(C))$ 
(Tab.~\ref{largetable}, bottom row). 

\begin{finding}
Hierarchy of bio-rationality of measures: 
Randi\'{c} index $>_{\beta}$ $\Pi$ $>_{\beta}$  Average edge length, Harary index $>_{\beta}$ 
 Average degree, Number of independent cycles  $>_{\beta}$ shortest path, diameter 
$>_{\beta}$ average cohesion  $>_{\beta}$ diameter in nodes $>_{\beta}$ shortest path in nodes.
\end{finding}

Matching between motorway graphs and Physarum graphs is most strongly expressed in the Randi\'{c} index; further measures amongst the top ones
are $\Pi$-index, Harary index, and edge length. Thus we can enhance the Finding~\ref{randic} as follows. 

\begin{finding}
The Randi\'{c} index is the most bio-compatible measure of transport networks.
\end{finding}

The Randi\'{c} index is used to characterise relations between structure, property and activity of chemical molecules~\cite{estrada_2001}, 
thus we can speculate that in terms of structure-property-activity Physarum almost perfectly approximates motorway networks in all regions!

 \subsubsection{Bio-rationality of motorways}


We calculate a bio-rationality of a motorway graph as follows 
$\rho(C) =\sum_{\mu}| \xi(|1 - \frac{\mu(\mathbf{H}(C))}{\mu(\mathbf{P}(C))}| \leq \epsilon)$, $\xi(p)=1$ if predicate $p$ is true, 
and 0 otherwise. We chose $\epsilon=0.1$. Values of bio-rationality of motorways are shown in last column in Tab.~\ref{largetable}.

\begin{finding}
Hierarchy of bio-rationality of regions is as follows: \\
\{Belgium, Canada\} $>_\rho$ China $>_\rho$ \{ Italy, Malaysia \}  $>_\rho$ the Netherlands  \\
$>_\rho$ \{ Brazil, Germany, Mexico, UK \} $>_\rho$ \{ Africa, USA \}
\label{biorationality}
\end{finding}

Comparing hierarchies of absolute matching, see Finding~\ref{findingAbsoluteMatching}, with the above hierarchy of bio-rationality we find 
Belgium, Canada, China and Italy are at the intersection of first three levels of the hierarchies. We omit Italy because its shape is intrinsically 
restrictive and invokes rather trivial architectures of protoplasmic networks. 

\begin{finding}
Motorway networks in Belgium, Canada and China are most affine to protoplasmic networks of slime mould P. polycephalum.
\end{finding}

\section{Discussion}

Based on the results of our previous laboratory experiments with slime mould imitating the development of transport networks in fourteen regions~\cite{PhysarumIberia}--\cite{PhysarumChina},\cite{PhysarumItaly} we undertook a comparative analysis of the motorway and protoplasmic networks. We found that  in terms of absolute matching between slime mould networks and motorway networks the regions studied can be arranged in the following 
order of decreasing matching: Malaysia, Italy, Canada, Belgium, China, Africa, the Netherlands, Germany, UK, Australia, Iberia, Mexico, Brazil, USA.  We compared the Physarum and the motorway graphs using such measures as average and longest shortest paths, average degrees, number of independent cycles, the Harary index, the $\Pi$-index and the Randi\'{c} index. We found that in terms of these measures motorway networks in Belgium, Canada and China are most affine to protoplasmic networks of slime mould \emph{P. polycephalum}. With regards to measures and topological indices we demonstrated that  the Randi\'{c} index could be considered as most bio-compatible measure of transport networks, because it matches incredibly well the slime mould and man-made transport networks, yet efficiently discriminates between transport networks of different regions.

\end{document}